\documentclass[aps,floatfix,amsmath,nofootinbib,amssymb,superscriptaddress]{revtex4}

\usepackage{overpic}
\usepackage{amssymb}
\usepackage{indentfirst}
\usepackage{feynmf}   %{feynmp}
\usepackage{slashed}  %for Feynman symbols
\usepackage{cases}
\usepackage{color}
\usepackage{multirow}
\usepackage{epstopdf}
\usepackage{graphicx,color,bm}
\usepackage{epstopdf}

\usepackage[colorlinks,
            citecolor=blue,
            anchorcolor=red,
            menucolor=red,
            linkcolor=red,
            filecolor=red,
            runcolor=red,
            urlcolor=blue,
            frenchlinks=red]{hyperref}

\begin{document}

\title{The transverse polarization of $\Lambda$ hyperons in $e^+e^-\to\Lambda^\uparrow h X$ processes within TMD factorization}
\author{Hui Li}
\affiliation{School of Physics, Southeast University, Nanjing 211189, China}
\author{Xiaoyu Wang}
\email{xiaoyuwang@zzu.edu.cn}
\affiliation{School of Physics and Microelectronics, Zhengzhou University, Zhengzhou 450001, China}
\author{Yongliang Yang}
\affiliation{College of Physics, Qingdao University, Qingdao 266071, China}
\author{Zhun Lu}
\email{zhunlu@seu.edu.cn}
\affiliation{School of Physics, Southeast University, Nanjing 211189, China}

\begin{abstract}

We investigate the transverse polarization of the $\Lambda$ hyperon in the processes $e^+e^-\to\Lambda^\uparrow \pi^\pm X$ and $e^+e^-\to\Lambda^\uparrow K^\pm X$ within the framework of the transverse momentum dependent~(TMD) factorization.
The transverse polarization is contributed by the convolution of the transversely polarizing fragmentation function~(PFF) $D_{1T}^\perp$ of the lambda hyperon and the unpolarized fragmentation function $D_1$ of pion/kaon.
We adopt the spectator diquark model result for $D_{1T}^{\perp}$ to numerically estimate the transverse polarization in $e^+e^-\to\Lambda^\uparrow h X$ process at the kinematical region of Belle Collaboration.
To implement the TMD evolution formalism of the fragmentation functions, we apply two different parametrizations on the nonperturbative Sudakov form factors associated with the fragmentation functions of the $\Lambda$, pion and kaon. It is found that our prediction on the polarization in the $\Lambda \pi^+$ production and $\bar{\Lambda} \pi^-$ is consistent with the recent Belle measurement in size and sign, while the model predictions on the polarizations in $\Lambda\pi^-$ and $\Lambda K^\pm$ productions show strong disagreement with the Belle data.
The reason for the discrepancies is discussed and possible approaches to improve the calculation in the future are also discussed.
\end{abstract}

\maketitle

\section{Introduction}

Understanding the internal parton structure of hadrons and the fragmentation mechanism of the parton into hadrons is one of the main goals in QCD and high energy physics.
Particularly, the production of a polarized $\Lambda$ hyperon from unpolarized $pp$ collisions has been observed~\cite{Lesnik:1975my,Bunce:1976yb} and it formed a long-standing challenge~\cite{Kane:1978nd,Dharmaratna:1996xd} in hadron physics and spin physics.
It is suggested~\cite{Anselmino:2000vs} that a polarizing fragmentation function~(PFF)~\cite{Mulders:1995dh}, denoted by $D_{1T}^\perp$, can account for the polarization of the $\Lambda$ production.
As a time-reversal-odd and transverse momentum dependent (TMD) fragmentation function, $D_{1T}^\perp$ describes the fragmentation of an unpolarized quark to a transversely polarized hadron, and it is usually viewed as the analog of the Sivers function~\cite{Sivers:1989cc,Sivers:1990fh} which gives the azimuthal asymmetry in the distribution of unpolarized quarks inside a transversely polarized nucleon.
Furthermore, $D_{1T}^\perp$ may play an important role in the spontaneous polarization, such as: $q\rightarrow \Lambda^{\uparrow}X$~\cite{Boer:2009uc}.
Thus, the study of the polarized $\Lambda$ production could also provide the information on the spin structure of the hyperon. This is intriguing since the $\Lambda$ hyperon can not serve as a target in high energy scattering processes.

As $D_{1T}^\perp$ is a chiral-even function, in principle it may be accessed directly without any unknown, chiral-odd, counterparts.
However, significant signal on the transverse polarization of the $\Lambda$ hyperon has not been observed in the single inclusive $e^+\,e^-$ annihilation (SIA) experiment performed by OPAL at LEP~\cite{Ackerstaff:1997nh}.
Alternative to SIA, $e^+e^-\rightarrow \Lambda^\uparrow+h+X$~\cite{Boer:1997mf,Wei:2014pma,Guan:2018ckx}  and semi-inclusive deep inelastic scattering (SIDIS) $\ell\,p\rightarrow\ell'+\Lambda^\uparrow+X$ were also suggested~\cite{Boer:1997mf} to study the $\Lambda$ polarization.
Those measurements could provide further understanding of the origin of the sizable transverse polarisation of hyperons observed in different  processes~\cite{Anselmino:2000vs,Mulders:1995dh,Bunce:1976yb,Heller:1978ty,Anselmino:2001js,
Koike:2017fxr,Gamberg:2018fwy}.
Recently, nonzero transverse polarization of $\Lambda$ production in SIA and semi-inclusive $e^+e^-\rightarrow\Lambda(\bar{\Lambda})+K^{\pm}(\pi^{\pm})+X$ process was measured by the Belle Collaboration~\cite{Guan:2018ckx}, making the extraction~\cite{DAlesio:2020wjq,Callos:2020qtu} of the polarized fragmentation function of the $\Lambda$ possible.
It is worth pointing out that, since there are no hadrons in the initial state, electron-positron annihilation is recognized as a rather clean process to access TMD fragmentation functions.
Motivated by the recent Belle data, the authors in Ref.~\cite{Anselmino:2019cqd} applied an approach that assumes a kinematical configuration between the PFF and the unpolarized fragmentation function, which allows the comparison between numerical calculations and experimental data in the SIA process.
The comparison showed the estimate results are in agreement with the data.
On the other hand, model calculations may also provide an approach to acquire knowledge of the transverse polarization of the $\Lambda$ hyperon.
A calculation of $D_{1T}^\perp$ for light flavors based on a spectator-diquark model has been performed in Ref.~\cite{Yang:2017cwi} and was used to make predictions on physical observables.

The main purpose of this work is to apply the TMD factorization ($P_{hT}/z_h\ll Q$)~\cite{Collins:1981uk,Collins:1984kg,Ji:2004wu,Ji:2004xq,Collins:2011zzd} to estimate the transverse polarization in the transversely polarized $\Lambda$ production $e^+e^-\rightarrow\Lambda(\bar{\Lambda})+\pi^{\pm}(K^{\pm})+X$ process.
In the last decades, TMD factorization has been applied in various high energy processes~\cite{Boer:2008fr,Arnold:2008kf,Aybat:2011zv,Collins:2011zzd,Collins:2012uy,Echevarria:2012pw,
Pitonyak:2013dsu,Echevarria:2014xaa,Kang:2015msa,Bacchetta:2017gcc,Wang:2017zym}.
In the TMD formalism, the differential cross section in the region $P_{hT}/z_h \ll Q$ can be expressed as the convolution of the hard scattering factors and the well-defined TMD distributions or fragmentation functions.
The TMD formalism embeds evolution information of those functions, of which the energy evolution (or the scale dependence) are governed by the so-called Collins-Soper equation~\cite{Collins:1981uk,Collins:1984kg,Collins:2011zzd,Idilbi:2004vb}.
The solution of the evolution equation indicates that the changes of TMDs from a initial scale to another scale may be determined by an exponential form of the Sudakov-like form factor~\cite{Collins:1984kg,Collins:2011zzd,Aybat:2011zv,Collins:1999dz}, which can be separated to the perturbative part and nonperturbative part.
The former one is perturbatively calculable, while the later one can not be calculated directly and is usually obtained by phenomenological extraction from experimental data.
In Refs.~\cite{Aybat:2011zv,Aybat:2011ge,Echevarria:2014xaa,Bacchetta:2017gcc} the authors extracted the nonperturbative Sudakov form factor corresponding to the unpolarized fragmentation function.
In the $e^+e^-\to\Lambda^\uparrow h^\pm X$ process, the convolution of the PFF for the $\Lambda$ hyperon and the unpolarized fragmentation function for pion/kaon can give rise to the transverse polarization of $\Lambda$ hyperon.
Particularly, we will take into account the TMD evolution for both the PFF $D_{1T}^\perp$ and the unpolarized fragmentation function $D_1$.
As a comparison, we adopt two different parameterizations on the nonperturbative part for the fragmentation functions~\cite{Echevarria:2014xaa,Bacchetta:2017gcc}.

The remaining content of the paper is organized as follows.
In Sec.~\ref{Sec.formalism}, we present the formalism of the transverse polarization contributed by the convolution of $D_{1T}^\perp$ and the unpolarized fragmentation function $D_1$ in the $e^+e^-\rightarrow \Lambda^\uparrow+h+X$ process within TMD factorization.
In Sec.~\ref{Sec.evolution}, we investigate the evolution effect for the unpolarized and transversely polarized fragmentation functions at leading order and present our choice on the nonperturbative Sudakov form factors associated with the fragmentation functions in details.
In Sec.~\ref{Sec.numerical}, we numerically estimate the transverse polarization at the energy $\sqrt{s}=10.58$ GeV which is accessible at Belle.
We also compare the results calculated from different choices of the nonperturbative ingredients associated with the TMD evolution.
Finally, We summarize the paper in Sec.~\ref{Sec.conclusion}.

\section{transverse polarization in $e^+e^-\rightarrow \Lambda^\uparrow+h+X$ process}

\label{Sec.formalism}
In this section, we will present the detailed framework of transverse polarization in $e^+e^-$ annihilation process with a transversely polarized $\Lambda$ hyperon and a light meson $h$~(pion or kaon) produced in the final state
\begin{align}
e^+(\ell)+e^-(\ell^\prime)\rightarrow \gamma^*(q)+X\rightarrow q(k)+\bar{q}(p)+X\rightarrow \Lambda^\uparrow(K)+h(P)+X.
\end{align}
In the $e^+e^-$ annihilation process, the electron and the positron annihilate into a virtual photon, which then decays into a quark-antiquark pair, the latter one fragments into final-state transversely polarized $\Lambda$ hyperon and the light meson $h$~(pion or kaon in this work), respectively.
$\ell$ and $\ell^\prime$ stand for the four-momenta of the initial-state positron and electron, $q$ is the four-momentum of the virtual photon with $q=\ell+\ell^\prime$, which is time-like~($q^2=Q^2>0$).
$k$ and $p$ are the four-momenta of the quark and antiquark, respectively, while $K$ and $P$ are the four-momenta of the final-state $\Lambda$ and pion/kaon meson.
The center of mass energy for the process can be written as $s=(\ell+\ell^\prime)^2=Q^2$.

In the ideal case, the transversely polarized $\Lambda$ and the pion/kaon meson should be produced completely back-to-back. However, the radiation of the gluon in the fragmentation process and the transverse momentum dependence make the hadrons deviate from the ideal back-to-back state.
The TMD factorization can be used to describe the imbalance from the back-to-back state as well as to calculate the differential cross section.
There are two experimental methods to define the reference frame in the $e^+e^-$ annihilation process in literature~\cite{Seidl:2008xc,Anselmino:2008jk,Anselmino:2015sxa,Boer:2008fr}.
In this work, we adopt the second-hadron momentum frame, which means that the momentum direction of the second hadron-pion/kaon meson is defined as the $z$ axis.
The $\hat{xz}$ plane is determined by the lepton and the pion/kaon meson momentum directions, while the hadron plane is determined by $z$ axis and the momentum direction of the $\Lambda$ hadron.
Hence $\phi$ is the azimuthal angle of the hadron plane relative to the lepton plane, while $\phi_S$ represents the azimuthal angle of the $\Lambda$ hyperon polarization vector $S_\perp$ in the lepton frame. $\bm{k}_T$ and $\bm{p}_T$ are the transverse momenta of the two fragmenting quarks, which are related to the transverse momenta of the final hadrons through $\bm{K}_\perp=-z_1\bm{k}_T$ and $\bm{P}_\perp=-z_2\bm{p}_T$.
Finally, $\bm{P}_{\Lambda\perp}=-z_1\bm{q}_T$ is the transverse momentum of $\Lambda$ hadron with respect to the transverse momentum of the virtual photon $\bm{q}_T$, which is also recognised as the transverse momentum of the process.
The invariants $z_1$ and $z_2$ are defined as $z_1=\frac{2K\cdot q}{Q^2}$ and $z_2=\frac{2P\cdot q}{Q^2}$, which can be identified as the light-cone momentum fraction (neglecting term of order ($q_T^2/(Q^2))$) of the $\Lambda$ and the pion/kaon meson in the fragmenting quark and antiquark, respectively.

The transverse polarization $P_{\Lambda T}$ in $e^+ e^- \to \Lambda^\uparrow h X$ can be defined as~\cite{Anselmino:2019cqd,Callos:2020qtu}
\begin{align}
P_{\Lambda T}=\frac{d\Delta \sigma}{d\sigma}=\frac{\mathcal{F}[\bm{\hat{h}}\cdot\bm{k}_T \frac{D_{1T}^{\perp}\bar{D}_1}{M_\Lambda}]}{2\mathcal{F}[D_1\bar{D}_1]},
\label{polar}
\end{align}
where $d\Delta \sigma=\frac{1}{2}(\sigma(S_\perp)-\sigma(-S_\perp))$, and $d\sigma$ in the denominator is the unpolarized differential cross section.
The transverse-spin dependent differential cross section can be expressed as~\cite{Callos:2020qtu}
\begin{align}
&\frac{d\sigma(S_\perp)}{dz_1dz_2 d(\cos\theta) d^2q_T}=\frac{N_c\pi\alpha_{em}^2}{2Q^2}(1+\cos^2\theta)z_1^2z_2^2
\left\{\mathcal{F}[D_1\bar{D}_1]
+|S_{\perp}|\sin(\phi_S-\phi)\mathcal{F}[\bm{\hat{h}}\cdot \bm{k}_T \frac{D_{1T}^{\perp}\bar{D}_1}{M_\Lambda}]\right\},
\end{align}
in which the azimuthal angle of the lepton plane $\phi^l$ has been integrated out.
Here, $M_\Lambda$ is the mass of the $\Lambda$, and the unit vector $\bm{\hat{h}}$ is defined as $\bm{\hat{h}}=\frac{\bm{P}_{\Lambda\perp}}{|\bm{P}_{\Lambda\perp}|} = \frac{\bm{q}_{T}}{q_T}$~\cite{Boer:1999mm,Arnold:2008kf}.
The notation $\mathcal{F}$ represents the convolution of the corresponding fragmentation functions in the transverse momentum space
\begin{align}
&\mathcal{F}[\omega D\bar{D}]= \sum_{q}e_q^2\int d^2\bm{k}_Td^2\bm{p}_T\delta^2(\bm{q}_T-\bm{k}_T-\bm{p}_T)
\omega(\bm{p}_T,\bm{k}_T)D^{\Lambda/q}(z_1,z_1^2\bm{k}_T^2)\bar{D}^{h/\bar{q}}(z_2,z_2^2\bm{p}_T^2),
\end{align}
with $\omega(\bm{p}_T,\bm{k}_T)$ an arbitrary function of $\bm{p}_T$ and $\bm{k}_T$.
Since it is convenient to deal with the TMD evolution effect in the $b$ space, which is conjugated to the $k_T$ space, we perform the Fourier transformation for the delta function
\begin{align}
\delta^{2}(\bm{q}_T-\bm{k}_T-\bm{p}_T) = {1\over (2\pi)^2}\int d^2 \bm{b}_\perp e^{-i \bm b_\perp\cdot(
\bm{q}_T-\bm{k}_T-\bm{p}_T)}
\end{align}
to obtain the denominator in Eq.~(\ref{polar}) as
\begin{align}
\mathcal{F}[D_1\bar{D}_1]
=&\sum_q e_q^2 \int{d^2\bm{k}_T d^2\bm{p}_T \delta^2(\bm{q}_T-\bm{k}_T-\bm{p}_T)D_1^{\Lambda/q}(z_1,z_1^2\bm{k}_T^2;Q)
\bar{D}_1^{h/\bar{q}}(z_2,z_2^2\bm{p}_T^2;Q)}\nonumber\\
=&\frac{1}{z_1^2}\frac{1}{z_2^2}\sum_{q}e_q^2\int d^2\bm{K}_\perp d^2\bm{P}_\perp\delta^2(-{\bm{P}_{\Lambda\perp}}/{z_1}+{\bm{K}_\perp}/{z_1}+{\bm{P}_\perp}/{z_2})
D_1^{\Lambda/q}(z_1,\bm{K}_\perp^2;Q)\bar{D}_1^{h/\bar{q}}(z_2,\bm{P}_\perp^2;Q)\nonumber\\
=&\frac{1}{z_1^2}\frac{1}{z_2^2}\sum_{q}e_q^2 \int\frac{d^2\bm{b}_\perp}{(2\pi)^2}e^{i{\bm{P}_{\Lambda\perp}}\cdot\bm{b}_\perp/{z_1}}
\tilde{D}_1^{\Lambda/q}(z_1,b;Q)\tilde{\bar{D}}_1^{h/\bar{q}}(z_2,b;Q)\nonumber\\
=&\frac{1}{z_1^2}\frac{1}{z_2^2}\sum_{q}e_q^2 \int_0^\infty\frac{dbb}{(2\pi)}J_0(P_{\Lambda\perp} b/z_1)
\tilde{D}_1^{\Lambda/q}(z_1,b;Q)\tilde{\bar{D}}_1^{h/\bar{q}}(z_2,b;Q),
 \label{eq:FUU}
\end{align}
where $J_0$ is the Bessel function of the zeroth order.
Thus, the unpolarized fragmentation function in $b$ space~(hereafter the tilde terms represent the ones in $b$ space, with b=$|{\bm{b}_\perp}|$) is the Fourier transformation of the fragmentation function in momentum space $D_1(z,\bm{P}_\perp^2;Q)$
\begin{align}
\tilde{D}_1(z,b;Q)&=\int{d^2\bm{P}_{\perp} e^{-i\bm{P}_{\perp} \cdot \bm{b}_\perp/z} D_1(z,\bm{P}_{\perp}^2;Q)},
\label{eq:UPFF}
\end{align}
with $\bm{P}_\perp$ the transverse momentum of the hadron with respect to the fragmenting quark, and
$P_\perp=|\bm{P}_\perp|$.
Similarly, the numerator can be expressed as:
\begin{align}
&\mathcal{F}[\bm{\hat{h}}\cdot\bm{k}_T \frac{D_{1T}^{\perp}\bar{D}_1}{M_\Lambda}]\nonumber\\
=&\frac{1}{z_1^2}\frac{1}{z_2^2}\sum_{q}e_q^2\int d^2\bm{K}_\perp d^2\bm{P}_\perp\delta^2(-{\bm{P}_{\Lambda\perp}}/{z_1}+{\bm{K}_\perp}/{z_1}+{\bm{P}_\perp}/{z_2})
(-\frac{\bm{\hat{h}}\cdot\bm{K}_\perp}{z_1M_\Lambda})D_{1T}^{\perp\Lambda^\uparrow/q}(z_1,\bm{K}_\perp^2;Q)
\bar{D}_1^{h/\bar{q}}(z_2,\bm{P}_\perp^2;Q)\nonumber\\
=&-\frac{1}{z_1^3}\frac{1}{z_2^2}\sum_{q}e_q^2 \int\frac{d^2\bm{b}_\perp}{(2\pi)^2}e^{i{\bm{P}_{\Lambda\perp}}\cdot\bm{b}_\perp/{z_1}}
\bm{\hat{h}}_\alpha
\tilde{D}_{1T}^{\perp\Lambda^\uparrow/q(\alpha)}(z_1,b;Q)\tilde{\bar{D}}_1^{h/\bar{q}}(z_2,b;Q),
\label{eq:FUT}
\end{align}
where the PFF of the $\Lambda$ hyperon in $b$ space is defined as
\begin{align}
\tilde{D}_{1T}^{\perp\Lambda^\uparrow/q(\alpha)}(z,b;Q)&=\int{d^2\bm{K}_\perp e^{-i\bm{K}_\perp\cdot \bm{b}_\perp/z} \frac{\bm{K}_{\perp\alpha}}{M_\Lambda} D_{1T}^{\perp\Lambda^\uparrow/q}(z,\bm{K}_\perp ^2;Q)}
\label{eq:PFF}
\end{align}
The energy dependence of the fragmentation functions will be discussed in details in the following section.

\section{The TMD evolution of fragmentation functions}
\label{Sec.evolution}

In this section, we will set up the formalism of the TMD evolution for both the PFF and the unpolarized fragmentation function.

The key purpose of the TMD evolution is to deal with the energy dependence of the TMD fragmentation functions in the $b$ space.
The advantage of the $b$ space is that the differential cross section can be written as a simple product instead of complicated convolution of the fragmentation functions in the transverse momentum space.
There are two energy dependencies of the TMD fragmentation functions $\tilde{D}(z,b;\mu,\zeta_D)$ in $b$ space, one is the renormalization scale $\mu$ related to the collinear fragmentation functions, and the other one $\zeta_D$ is the scale related to the cutoff in the operator definition of the TMD function to regularize the singularity~\cite{Collins:1981uk,Collins:1984kg,Collins:2011zzd,Aybat:2011zv,Aybat:2011ge,Echevarria:2012pw}.
Hereafter, we set $\mu=\sqrt{\zeta_D}=Q$ for simplicity, and the TMD fragmentation functions can be written as $\tilde{D}(z,b;Q)$~(Here, $D$ is a shorthand for any fragmentation function, such as the PFF $D_{1T}^\perp$ and the unpolarized fragmentation function $D_1$).
Therefore, the main focus turns to solve the energy evolution equations of $\tilde{D}(z,b;Q)$
~\cite{Collins:2011zzd,Echevarria:2012js} defined in Eqs.~(\ref{eq:UPFF}) and (\ref{eq:PFF}).

\subsection{Solving the TMD evolution equation for the fragmentation function in $b$-space}

The energy dependence for the $\zeta_D$ is encoded in the Collins-Soper-Sterman (CSS) equation
\begin{align}
\frac{\partial\ \mathrm{ln} \tilde{D}(z,b;\mu,\zeta_D)}{\partial\ \sqrt{\zeta_D}}=\tilde{K}(b;\mu),
\end{align}
while the $\mu$ dependence is derived from the renormalization group equation as
\begin{align}
&\frac{d\ \tilde{K}}{d\ \mathrm{ln}\mu}=-\gamma_K(\alpha_s(\mu)),\\
&\frac{d\ \mathrm{ln} \tilde{D}(z,b;\mu,\zeta_D)}
{d\ \mathrm{ln}\mu}=\gamma_D(\alpha_s(\mu);{\frac{\zeta^2_D}{\mu^2}}),
\end{align}
with $\tilde{K}$ the evolution kernel, and $\gamma_K$, $\gamma_D$ the anomalous dimensions.
The solutions of these evolution equations were studied in details in Refs.~\cite{Idilbi:2004vb,Collins:2011zzd,Collins:2014jpa}.
Here, we will only discuss the final result.
After solving the above evolution equations, the overall structure of the solutions are identical to each other in all TMD factorization schemes, and the evolution effects are incorporated into the exponential form factors~\cite{Collins:1981uk,Collins:1984kg,Collins:2011zzd,Ji:2004wu,Collins:2014jpa,Ji:2004xq} as
\begin{equation}
\tilde{D}(z,b;Q)=\mathcal{D}(Q)\times e^{-S(Q,b)}\times \tilde{D}(z,b,\mu_i).
\label{eq:Sudakov factor}
\end{equation}
Here, $\mathcal{D}(Q)$ is the hard scattering factor which can be calculated through perturbative QCD, and $S(Q,b)$ is the Sudakov-like form factor.
Eq.~(\ref{eq:Sudakov factor}) shows that the TMD fragmentation functions $D$ at an arbitrary scale $Q$ can be evolved from an initial scale $\mu_i$ through the evolution encoded by the exponential form $\exp(-S(Q,b))$.

Studying the behavior of the TMD fragmentation function in the $b$ space is quite important since it can determine the transverse momentum dependence of the physical observables through the inverse Fourier transformation.
To do this, one need the information of $S(Q,b)$ in the entire $b$ region.
In the small $b$ region $1/Q \ll b \ll 1/ \Lambda$, the $b$-dependence is perturbative, while it turns to be non-perturbative in the large $b$ region.
Thus, to combine the information between the two regions, a matching procedure should be adopted.
In the original CSS approach~\cite{Collins:1984kg,Qiu:2000ga,Qiu:2000hf,Landry:2002ix},
a parameter $b_{\mathrm{max}}$ is introduced as the boundary between the two regions, which allows a smooth transition from perturbative region to nonperturbative region as well as to avoid hitting on the Landau pole.
A $b$-dependent function $b_\ast(b)$ may be also introduced to have the property $b_\ast\approx b$ at small $b$ value and $b_{\ast}\approx b_{\mathrm{max}}$ at large $b$ value.
The typical value of $b_{\mathrm{max}}$ is chosen around $1\ \mathrm{GeV}^{-1}$ to guarantee that $b_{\ast}(b)$ is always in the perturbative region.
In the original CSS approach it has the following form~\cite{Collins:1984kg,Kang:2015msa}
\begin{align}
\label{eq:b*}
b_\ast=b/\sqrt{1+b^2/b_{\rm max}^2}  \ ,~b_{\rm max}<1/\Lambda_\mathrm{QCD}.
\end{align}
Apart from the above form, there are also several different choices on the form of $b_\ast(b)$ in literature~\cite{Collins:2016hqq,Bacchetta:2017gcc}.

\subsection{Sudakov form factor}

After introducing the $b_{\rm max}$ and the $b_{\ast}$ prescription, the Sudakov form factor can be separated into two parts: the perturbative part $S_\mathrm{P}$ and the non-perturbative part $S_\mathrm{NP}$.
It is important to keep in mind that $\tilde{D}_1^{h/q}(z,b;Q)$, $\tilde{D}_1^{\Lambda/q}(z,b;Q)$ and $D_{1T}^{\perp\Lambda/q(\alpha)}(z,b;Q)$ follow exactly the same QCD evolution effect in the perturbative region, which means that $S_{\mathrm{P}}(Q,b)$ is universal and is the same for different kinds of fragmentation functions (namely, $S_{P}$ is spin-independent).
The details of $S_\mathrm{P}$ have been studied in literature, it is usually written in the following general form~\cite{Echevarria:2014xaa,Kang:2011mr,Aybat:2011ge,Echevarria:2012pw,Echevarria:2014rua}:
\begin{equation}
\label{eq:Spert}
S_{\mathrm{P}}(Q,b)=\int^{Q^2}_{\mu_b^2}\frac{d\bar{\mu}^2}{\bar{\mu}^2}\left[A(\alpha_s(\bar{\mu}))
\mathrm{ln}\frac{Q^2}{\bar{\mu}^2}+B(\alpha_s(\bar{\mu}))\right],
\end{equation}
where the coefficients $A$ and $B$ in Eq.~(\ref{eq:Spert}) can be expanded as the series of $\alpha_s/{\pi}$:
\begin{align}
A=\sum_{n=1}^{\infty}A^{(n)}(\frac{\alpha_s}{\pi})^n,\\
B=\sum_{n=1}^{\infty}B^{(n)}(\frac{\alpha_s}{\pi})^n.
\end{align}
In this work, we adopt $A^{(n)}$ up to $A^{(2)}$ and $B^{(n)}$ up to $B^{(1)}$ in the accuracy of next-to-leading-logarithmic (NLL) order~\cite{Collins:1984kg,Landry:2002ix,Qiu:2000ga,Kang:2011mr,Aybat:2011zv,Echevarria:2012pw} :
\begin{align}
A^{(1)}&=C_F,\\
A^{(2)}&=\frac{C_F}{2}\left[C_A\left(\frac{67}{18}-\frac{\pi^2}{6}\right)-\frac{10}{9}T_Rn_f\right],\\
B^{(1)}&=-\frac{3}{2}C_F,
\end{align}
with $C_F=4/3$, $C_A=3$ and $T_R=1/2$.

On the other hand, the non-perturbative part of the Sudakov form factor $S_\mathrm{NP}$ can not be calculated perturbatively, it is usually parameterized and extracted from experimental data.
There are several different approaches to parameterize $S_\mathrm{NP}$, we will discuss two of them in details.

One of the widely used non-perturbative Sudakov form factor $S_\mathrm{NP}$ associated with an unpolarized TMDFFs has the following form (Echevarria-Idilbi-Kang-Vitev (EIKV) parametrization)~\cite{Landry:2002ix,Konychev:2005iy,Davies:1984sp,Ellis:1997sc}:
\begin{align}
\label{eq:SNP(unp-ff)}
S^{D_1}_\mathrm{NP}(b,Q)=b^2(g_1^\mathrm{ff}+\frac{g_2}{2}\ln{\frac{Q}{Q_0}}).
\end{align}
Since the information of the nonperturbative Sudakov form factor associated with the \textbf{polarized} fragmentation function of the $\Lambda$ hyperon still remains unknown, we assume it to be the same as the one for the unpolarized fragmentation function, i.e., $S^{D_{1T}^{\perp\Lambda/q}}_\mathrm{NP}=S^{D_1}_\mathrm{NP}$.

In Eq.~(\ref{eq:SNP(unp-ff)}), $g_i(b)$ are functions of the impact parameter $b$.
Particularly, $g_2(b)$ contains the information on the large $b$ behavior of the evolution kernel $\tilde K$, while $g_1^\mathrm{ff}$ contains the information about the intrinsic nonperturbative transverse motion of bound partons, i.e., it depends on the type of the hadron and quark flavor.
It might also depend on the momentum fraction of the hadrons in the fragmenting quark $z$~\cite{Su:2014wpa}.
One should note that $g_2(b)$ is universal among different types of TMDs and does not depend on the particular process, which is one of the important predictions of TMD factorization~\cite{Collins:2011zzd,Aybat:2011zv,Echevarria:2014xaa,Kang:2015msa}.
On the other hand, the parameter $g_1^\mathrm{ff}$ was assumed to be related to the intrinsic transverse momentum squared $\langle p_T^2\rangle_{Q_0}$ for TMD fragmentation functions at the initial energy scale $Q_0$ as
\begin{equation}
g_1^\mathrm{ff}=\frac{\langle p_T^2\rangle_{Q_0}}{4z^2}
\end{equation}
In Ref.~\cite{Echevarria:2014xaa}, the authors show that the Sudakov factor with the following parameters (and $Q_0=\sqrt{2.4}\ \textrm{GeV}$) leads to a reasonably good description of all experimental data on SIDIS, DY lepton pair and W/Z boson production,
\begin{equation}
\langle p_T^2\rangle_{Q_0}=0.19\ \textrm{GeV}^2, ~~~~~~g_2=0.16\ \textrm{GeV}^2,~~~~~b_\mathrm{max}=1.5\ \textrm{GeV}^{-1}.
\end{equation}

Besides the EIKV parametrization, several other forms for $S_\mathrm{NP}$ have been also proposed~\cite{Collins:2011zzd,Aybat:2011zv,Aybat:2011ge,Echevarria:2014xaa,Echevarria:2014rua,Kang:2011mr} recently.
Particularly, we will discuss the evolution formalism from Ref.~\cite{Bacchetta:2017gcc}, which is called as Bacchetta-Delcarro-Pisano-Radici-Signori (BDPRS) parametrization.
In this approach, the fragmentation functions evolving from the initial energy scale to the final energy scale also has the exponential form as
\begin{equation}
\label{eq:SNP-jhep1}
\tilde{D}^{a\rightarrow h}_1(z,b^2;Q^2)=D^{a\rightarrow h}_1(z;\mu_b^2)e^{-S(\mu_b^2,Q^2)}e^{\frac{1}{2}g_K(b)\ln(Q^2/Q_0^2)}\tilde{D}^{a\rightarrow h}_{1\mathrm{NP}}(z,b^2),
\end{equation}
where $g_K=-g_2b^2/2$, following the choice in Refs.~\cite{Landry:2002ix,Nadolsky:1999kb,Konychev:2005iy}. $\tilde{D}^{a\rightarrow h}_{1\mathrm{NP}}(z,b^2)$ is the intrinsic nonperturbative part of the fragmentation function parameterized as
\begin{equation}
\label{eq:SNP-jhep2}
\tilde{D}^{a\rightarrow h}_{1\mathrm{NP}}(z,b^2)=\frac{g_3e^{-g_3\frac{b^2}{4z^2}}
+(\frac{\lambda_F}{z^2})g_4^2(1-g_4\frac{b^2}{4z^2})e^{-g_4\frac{b^2}{4z^2}}}
{2\pi z^2 (g_3+(\frac{\lambda_F}{z^2})g_4^2)},
\end{equation}
with
\begin{equation}
g_{3,4}(z)=N_{3,4}\frac{(z^\beta+\delta)(1-z)^\gamma}{(\hat{z}^\beta+\delta)(1-\hat{z})^\gamma}
\end{equation}
$\beta, \gamma, \delta$ and $N_{3,4}\equiv g_{3,4}(\hat{z})$ with $\hat{z}=0.5$ are free parameters fitted to the available data from SIDIS, Drell-Yan, and Z boson production processes yielding $\beta=1.65, \gamma=2.28, \delta=0.14, \lambda_F=5.50\ \textrm{GeV}^{-2}, g_2=0.13\ \textrm{GeV}^{2}, N_3=0.21\ \textrm{GeV}^2, N_4=0.03\ \textrm{GeV}^2$.
Furthermore, in Ref.~\cite{Bacchetta:2017gcc}, the new $b_\ast$ prescription different from Eq.~(\ref{eq:b*}) was also proposed as
\begin{equation}
b_\ast=b_\mathrm{max}\left(\frac{1-e^{{-b^4}/{b_\mathrm{max}^4}}}{1-e^{{-b^4}/{b_\mathrm{min}^4}}}\right)^{1/4}
\label{b*2}
\end{equation}
Again, $b_\mathrm{max}$ is the boundary of the nonperturbative and perturbative $b$-space region fixed by $b_\mathrm{max}=2e^{-\gamma _E}\ \textrm{GeV}^{-1}\approx 1.123\ \textrm{GeV}^{-1}$, with $\gamma_E\approx0.577$ the Euler's constant~\cite{Collins:1981uk}.
Besides, the authors in Ref.~\cite{Bacchetta:2017gcc} also chose to saturate $b_\ast$ at the minimum value $b_\mathrm{min}\propto 2e^{-\gamma_E}/Q$.

In this work, we will adopt the both the EIKV evolution formalism and the BDPRS evolution formalism to calculate the transverse polarization of $\Lambda$ to investigate the impact of the different evolution formalisms on the polarization.

\subsection{The evolved fragmentation functions}

In the perturbative region $1/Q \ll b \ll 1/ \Lambda$, the TMD fragmentation functions can be expressed as the convolution of the perturbatively calculable coefficients and the corresponding collinear counterparts of the TMD fragmentation functions
\begin{equation}
\tilde{D}(z,b;\mu_b)=\sum_i \int_{z}^1\frac{d\xi}{\xi} C_{q\leftarrow i}(z/\xi,b;\mu)D_{i/H}(\xi,\mu).
\label{eq:small_b_F}
\end{equation}
at the fixed energy scale $\mu_b$,  which is a dynamic scale related to $b_\ast$ through $\mu_b=c/b_\ast$, with $c=2e^{-\gamma_E}$, %and $\gamma_E\approx0.577$ the Euler's constant~\cite{Collins:1981uk},
$C_{q\leftarrow i}(z/\xi,b;\mu)=\sum_{n=0}^{\infty}C_{q\leftarrow i}^{(n)}(\alpha_s/\pi)^n$ is the perturbatively calculable coefficient function with $\sum_i$ summing over the quark and antiquark flavors. Here, we will adopt the leading order (LO) result, i.e. $C_{q\leftarrow i}^{(0)}=\delta_{iq}\delta(1-z)$. In other words,
\begin{equation}
\tilde{D}_1^{h/q}(z,b;\mu_b)=D_1^{h/q}(z,\mu_b),
\label{C-d_1^pion}
\end{equation}
\begin{equation}
\tilde{D}_1^{\Lambda/q}(z,b;\mu_b)=D_1^{\Lambda/q}(z,\mu_b),
\label{C-d_1^lamda}
\end{equation}
\begin{equation}
\tilde{D}_{1T}^{\perp\Lambda/q(\alpha)}(z,b;\mu_b)=(\frac{ib^\alpha}{2})\hat{D}_{1T}^{\perp(3)}(z,z,\mu_b).
\label{C-d_1^tlamda}
\end{equation}
Here, $D_1^{h/q}(z,\mu_b)$ and $D_1^{\Lambda/q}(z,\mu_b)$ are the collinear unpolarized fragmentation functions for the pion/kaon meson and the $\Lambda$ hyperon, while $\hat{D}_{1T}^{\perp(3)}(z,z,\mu_b)$ is a twist-3 qqg correlation function related to the PFF $D_{1T}^{\perp(1)}$ ~\cite{Yuan:2009dw} as:
\begin{align}
\hat{D}_{1T}^{\perp(3)}(z,z,\mu_b)=\int{d^2\bm{K}_\perp\frac{|\bm{k}_\perp^2|}{M_\Lambda}D_{1T}^{\perp \Lambda/q}(z,\bm{K}_\perp^2)}=2M_{\Lambda}D_{1T}^{\perp(1)},
\label{twist-3 PFF}
\end{align}
where $D_{1T}^{\perp(1)}$ is the first transverse moment of $D_{1T}^{\perp\Lambda/q}$.

It is straightforward to rewrite the scale-dependent TMD fragmentation functions of the pion/kaon meson and the $\Lambda$ hyperon in $b$ space
\begin{align}
\label{eq:tildeF}
\tilde{D}_{h/q}(z,b;Q)=e^{-\frac{1}{2}S_{\mathrm{P}}(Q,b_\ast)-S^{D_{h/q}}_\mathrm{NP}(Q,b)}D_{h/j}(z,\mu_b),
\end{align}
The factor of $\frac{1}{2}$ in front of $S_{\mathrm{P}}$ comes from the fact that $S_{\mathrm{P}}$ of quarks and antiquarks satisfies the relation~\cite{Prokudin:2015ysa}
\begin{align}
S^q_{\mathrm{P}}(Q,b_\ast)=S^{\bar{q}}_{\mathrm{P}}(Q,b_\ast)=S_{\mathrm{P}}(Q,b_\ast)/2.
\end{align}
With all the above ingredients, we can explicitly write out the evolved TMD fragmentation functions as
\begin{align}
\tilde{D}_1^{h/q}(z,b;Q) &=e^{-\frac{1}{2}S_{\mathrm{P}}(Q,b_\ast)-S^{D_1^{h/q}}_{\mathrm{NP}}(Q,b)}
 D_1^{h/q}(z,\mu_b),\nonumber\\
 \tilde{D}_1^{\Lambda/q}(z,b;Q) &=e^{-\frac{1}{2}S_{\mathrm{P}}(Q,b_\ast)-S^{D_1^{\Lambda/q}}_{\mathrm{NP}}(Q,b)}
 D_1^{\Lambda/q}(z,\mu_b),\nonumber\\
\tilde{D}_{1T}^{\perp \Lambda/q(\alpha)}(z,b;Q) &=(\frac{ib^\alpha}{2})e^{-\frac{1}{2}S_{\mathrm{P}}(Q,b_\ast)-S^{{D}_{1T}^{\perp \Lambda/q}}_{\mathrm{NP}}(Q,b)}
 \hat{D}_{1T}^{\perp(3)}(z,z,\mu_b).
\label{eq:f_b}
\end{align}
Thus, the fragmentation function in the transverse momentum space can be obtained by performing the inverse Fourier transformation
\begin{align}
D_1^{h/q}(z,P_{\perp};Q)=\int_0^\infty\frac{db b}{2\pi}J_0(P_\perp b/z)e^{-\frac{1}{2}S_{P}(Q,b_\ast)-S^{D_1^{h/q}}_{\mathrm{NP}}(Q,b)} D_1^{h/q}(z,\mu_b),\label{eq:f-D1pion}\\
D_1^{\Lambda/q}(z,K_\perp;Q)=\int_0^\infty\frac{db b}{2\pi}J_0(K_\perp b/z)e^{-\frac{1}{2}S_{P}(Q,b_\ast)-S^{D_1^{\Lambda/q}}_{\mathrm{NP}}(Q,b)} D_1^{\Lambda/q}(z,\mu_b),\label{eq:f-D1Lambda}\\
\frac{K_{\perp\alpha}}{M_\Lambda}\tilde{D}_{1T}^{\perp \Lambda/q(\alpha)}(z,K_\perp;Q)=\int_0^\infty\frac{db b^2}{4\pi}J_1(K_\perp b/z)e^{-\frac{1}{2}S_{P}(Q,b_\ast)-S^{D_{1T}^{\perp \Lambda/q}}_{\mathrm{NP}}(Q,b)} D_{1T}^{\perp(3)}.
\end{align}
with $J_1$ being the Bessel function of the first order.

\section{Numerical calculation}

\label{Sec.numerical}

In this section, we numerically estimate the transverse polarization in the process $e^+e^-\rightarrow\Lambda(\bar{\Lambda})+K^{\pm}(\pi^{\pm})+X$ using the framework set up above and compare the estimation with recent experimental data measured by the Belle Collaboration~\cite{Guan:2018ckx}.

In order to obtain the numerical results, one needs to utilize the corresponding collinear parts of the TMD fragmentation functions as the inputs of the TMD evolution effects.
For the unpolarized collinear fragmentation function $D_1(z)$ for pion and kaon, we adopt the leading order set of DSS parametrization ~\cite{deFlorian:2007aj}.

\begin{figure}
  \centering
  % Requires \usepackage{graphicx}
  \includegraphics[width=0.32\columnwidth]{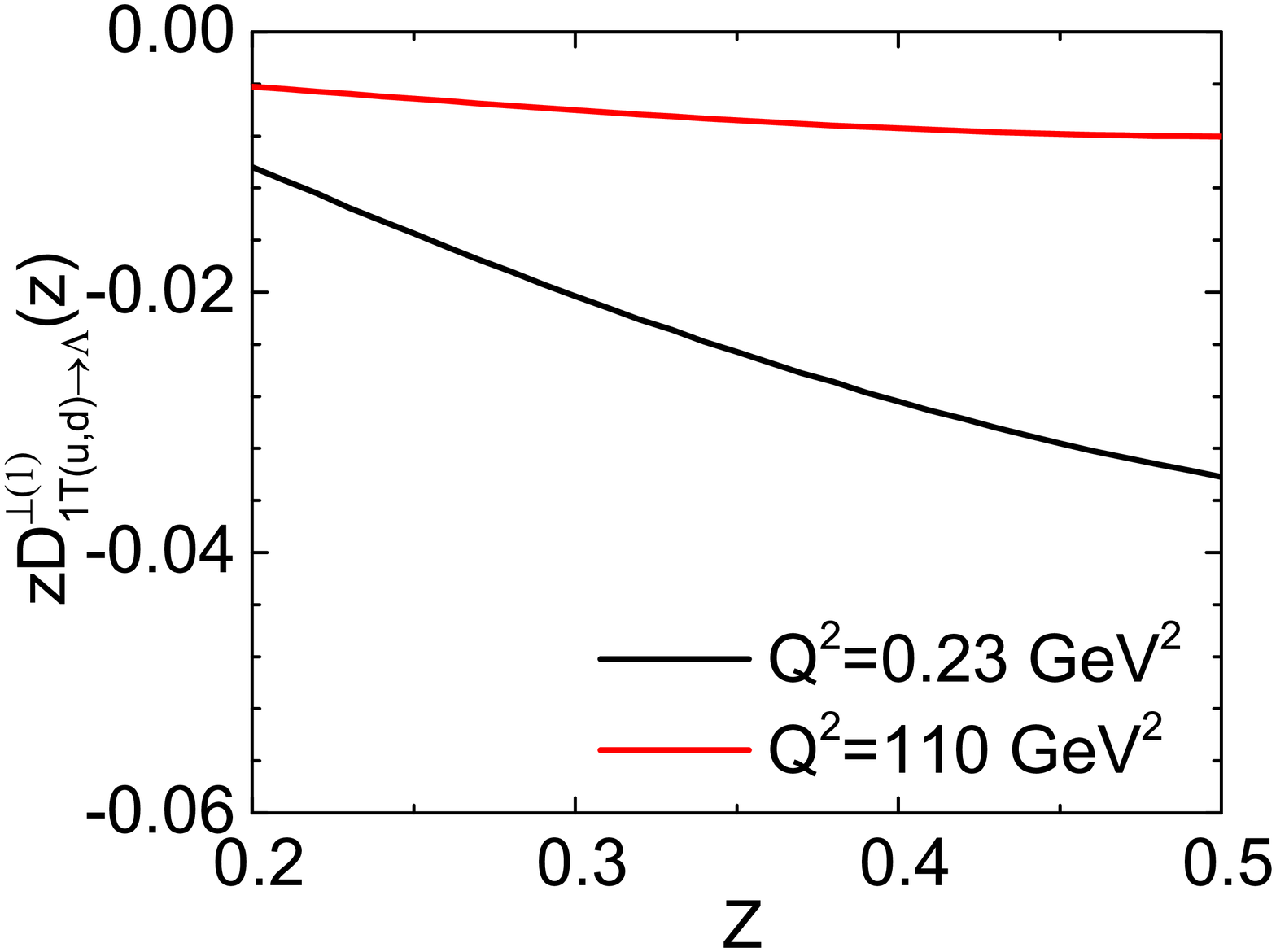}
  \includegraphics[width=0.32\columnwidth]{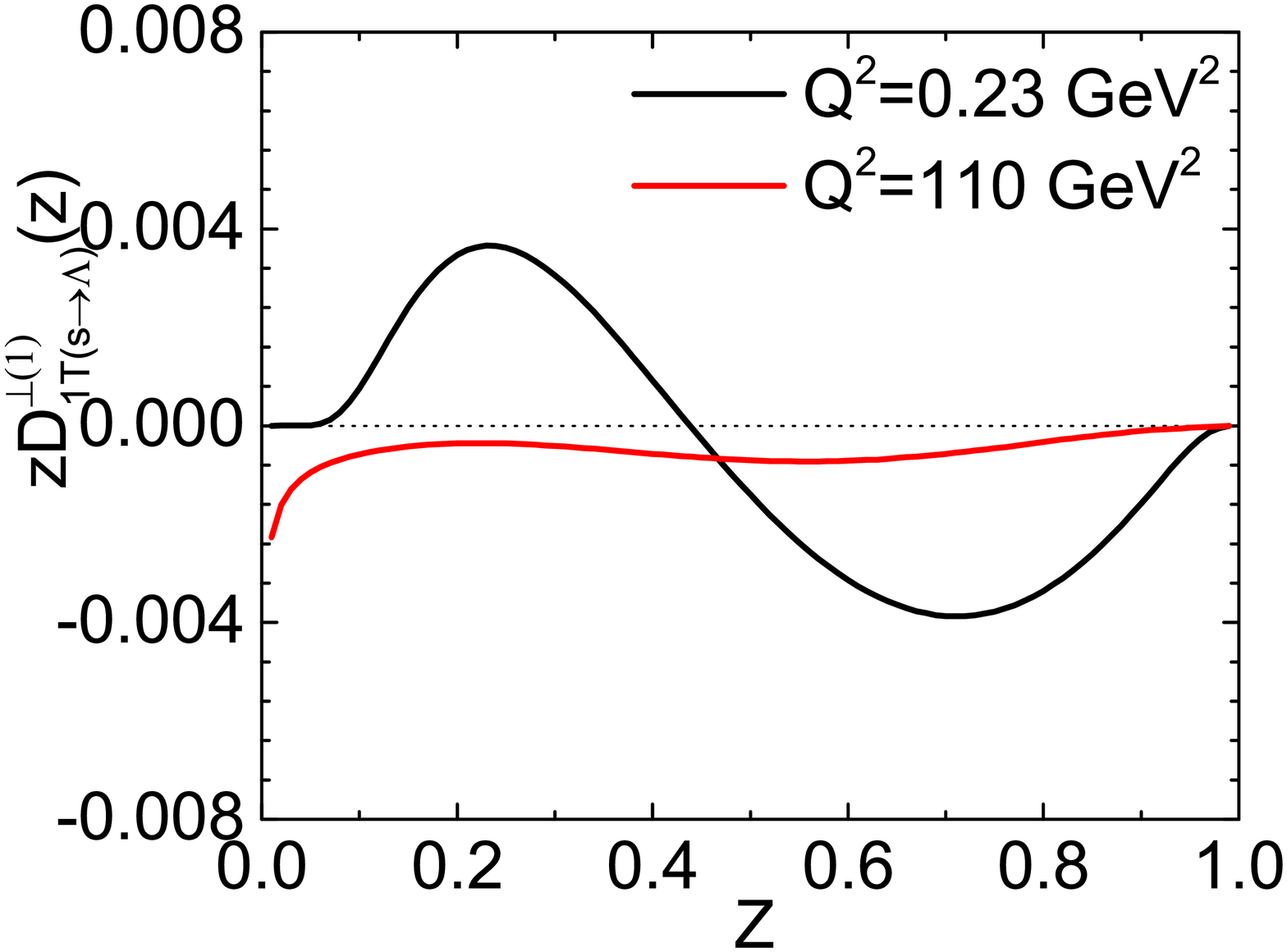}
  \caption{Left panel: the first transverse moment of  $D_{1T}^{\perp}$ (multiplied by $z$) of the up and down quark at $\mu_0^2=0.23$ GeV$^2$ and $Q^2$=110 GeV$^2$.
  Right panel: similar to the left panel, but for the strange quark.
  }
  \label{fig0}
\end{figure}

The other important input is the twist-3 collinear correlation function $D_{1T}^{\perp(3)}$, for which we apply the spectator diquark model results~\cite{Yang:2017cwi}, where the contributions from both the scalar diquark and the axial-vector diquark spectators are included.
Assuming the SU(6) spin-flavor symmetry, the fragmentation functions of the $\Lambda$ hyperon for light flavors satisfy the relations between different quark flavors and diquark types
\begin{align}
&D_{1T}^{\perp u}=D_{1T}^{\perp d}=\frac{1}{4}D_{1T}^{\perp (s)}+\frac{3}{4}D_{1T}^{\perp (v)},~~~~~~~~~D_{1T}^{\perp s}=D_{1T}^{\perp (s)}, \nonumber\\
&D_{1T}^{\perp \bar{u}} = D_{1T}^{\perp \bar{d}} =D_{1T}^{\perp \bar{s}}=0,
\label{PFF-uds}
\end{align}
where $u$, $d$ and $s$ denote the up, down and strange quarks,
respectively.
One should notice that in this model only the valence quarks contribute to the $\Lambda$ fragmentation function, while the sea quark contribution is zero.
$D_{1T}^{\perp (v)}$ and $D_{1T}^{\perp (s)}$ represent the contribution from the axial-vector diquark and scalar diquark, and have the form\footnote{The expression in Eq.~(\ref{axial vector diquark}) is the updated one from Eq.(35) of Ref.~\cite{Yang:2017cwi} by modifying the typos.}
\begin{align}
D_{1T}^{\perp(s)}(z,k_T^2)=&\frac{\alpha_s g_s^2 C_F}{(2\pi)^4}\frac{e^{-\frac{2k^2}{\Lambda^2}}}{z^2(1-z)}\frac{1}{(k^2-m_q^2)}\nonumber\\
&\times(D_{1T(a)}^{\perp(s)}(z,k_T^2)+D_{1T(b)}^{\perp(s)}(z,k_T^2)\nonumber\\
&+D_{1T(c)}^{\perp(s)}(z,k_T^2)+D_{1T(d)}^{\perp(s)}(z,k_T^2)),
\label{scalar diquark}\\
D_{1T}^{\perp(v)}(z,k_T^2)=&\frac{\alpha_s g_s^2 C_F}{(2\pi)^4}\frac{e^{-\frac{2k^2}{\Lambda^2}}}{z^2(1-z)}\frac{1}{(k^2-m_q^2)}\nonumber\\
&\times(D_{1T(a)}^{\perp(v)}(z,k_T^2)+D_{1T(b)}^{\perp(v)}(z,k_T^2)\nonumber\\
&+D_{1T(c)}^{\perp(v)}(z,k_T^2)+D_{1T(d)}^{\perp(v)}(z,k_T^2)),
\label{axial vector diquark}
\end{align}
where $k^2$ can be written as $k^2=\frac{z}{(1-z)}\bm{k}_T^2+\frac{m_D^2}{(1-z)}+\frac{M_\Lambda^2}{z}$, with $m_q, m_D, M_\Lambda$ the masses of the parent quark, the spectator diquark and fragmenting $\Lambda$ hyperon, respectively.
The $\Lambda^2$ has the general form $\Lambda^2=\lambda^2z^\alpha(1-z)^\beta$.
At one loop level, there are four diagrams that can
generate imaginary phases contributing to the nonzero polarized fragmentation function.
For the scalar case, the four nonzero contributions are
\begin{align}
D_{1T(a)}^{\perp(s)}(z,k_T^2)&=\frac{m_q M_\Lambda}{(k^2-m_q^2)}(3-\frac{m_q^2}{k^2})I_1,\nonumber\\
D_{1T(b)}^{\perp(s)}(z,k_T^2)&=M_\Lambda[m_q(2I_2-\mathcal{A})+M_\Lambda(2I_2-\mathcal{B}-2\mathcal{A})],\nonumber\\
D_{1T(c)}^{\perp(s)}(z,k_T^2)&=0,\nonumber\\
D_{1T(d)}^{\perp(s)}(z,k_T^2)&=\frac{M_\Lambda}{z}[2(1-z)(m_q \mathcal{C}P_\Lambda^--M_\Lambda\mathcal{D}P_\Lambda^-)+z(m_q A-M_\Lambda B)],
\end{align}
while for the axial-vector case, the four contributions are\footnote{The expression in Eq.~(\ref{diquark-abcd}) is the updated one from Eq.(36)-(39) of Ref.~\cite{Yang:2017cwi} by modifying the typos.}.
\begin{align}
D_{1T(a)}^{\perp(v)}(z,k_T^2)&=-\frac{m_q M_\Lambda}{(k^2-m_q^2)}(1-\frac{m_q^2}{3k^2})I_1,\nonumber\\
D_{1T(b)}^{\perp(v)}(z,k_T^2)&=\frac{1}{3}\{2M_\Lambda[m_q(I_2-\mathcal{A})+M_\Lambda(\mathcal{A}-I_2-\mathcal{B})]\nonumber\\
&+k.P_\Lambda(4I_2-6\mathcal{A})-\mathcal{A}k\cdot P_\Lambda-\mathcal{B}M_\Lambda^2\nonumber\\
&+\frac{3}{2}(\frac{k^2-m_q^2}{2k^2}I_1+(k^2-m_q^2)\mathcal{A})\},\nonumber\\
D_{1T(c)}^{\perp(v)}(z,k_T^2)&=0,\nonumber\\
D_{1T(d)}^{\perp(v)}(z,k_T^2)&=-\frac{1}{3M_\Lambda}\{[M_\Lambda((k^2-m_q^2)\mathcal{C}P_\Lambda^-+2M_\Lambda^2\mathcal{D}P_\Lambda^--2m_qM_\Lambda\mathcal{C}P_\Lambda^-)\nonumber\\
&+2k.P_\Lambda(m_q\mathcal{C}P_\Lambda^--M_\Lambda\mathcal{D}P_\Lambda^-)+z\frac{m_q}{2}I_1+\frac{k^2-m_q^2}{2}(M_\Lambda\mathcal{D}P_\Lambda^--m_q\mathcal{C}P_\Lambda^-)]\nonumber\\
&-M_\Lambda(m_qM_\Lambda\mathcal{A}+M_\Lambda^2\mathcal{B}+2k\cdot P_\Lambda\mathcal{A})-\frac{2M_\Lambda}{z}(m_qM_\Lambda\mathcal{C}P_\Lambda^-+k\cdot P_\Lambda\mathcal{C}P_\Lambda^-)
\}.
\label{diquark-abcd}
\end{align}
Here, the expressions for the functions of $\mathcal{A}, \mathcal{B}, \mathcal{C}P_\Lambda^-, \mathcal{D}P_\Lambda^-, I_i$ have been given in Ref.~\cite{Yang:2017cwi}.
Besides, the values of the model parameters are obtained by fitting the model result of the unpolarized fragmentation function $D_1^\Lambda$ to the DSV parametrization~\cite{deFlorian:1997zj} at the model scale $\mu_0^2=0.23\ \textrm{GeV}^2$ as
\begin{align}
m_D=0.745\ \textrm{GeV},\quad \lambda=5.967\ \textrm{GeV},\quad g_s=1.982\ \,\quad m_q=0.36\ \textrm{GeV},\quad M_\Lambda=1.116\ \textrm{GeV}, \quad \alpha=0.5,\quad \beta=0,
\label{parameter}
\end{align}
where the values of the last four parameters are fixed, and the coupling constant $\alpha_s$ is chosen as $0.817$ at the model scale.
Fig.~\ref{fig0} depicts the first transverse moment $D_{1T}^{\perp(1)}$ (multiplied by $z$) of the $\Lambda$ PFF  $D_{1T}^{\perp \Lambda/q}$ for light quark flavors.
The left panel shows the contribution for the $u$ and $d$ quarks, while the right panel shows the one for strange quark.
One can conclude that the size of $D_{1T}^{\perp(1)}$ for the up and down quarks is much larger than $D_{1T}^{\perp(1)}$ for the strange quark.

The collinear twist-3 fragmentation function of quark flavor $q$ to $\Lambda$ hyperon $\hat{D}_{1T}^{\perp(3)}(z,z,\mu_b)$ at the model scale can be obtained through Eq.~(\ref{twist-3 PFF}).
For consistency, we apply the unpolarized fragmentation function of the $\Lambda$ hyperon $D_1^{\Lambda/q}(z)$ calculated from the same model.
Furthermore, to solve the energy dependence of the collinear counterparts for the fragmentation functions, we apply the {\sc{QCDNUM}} evolution package~\cite{Botje:2010ay} to evolve the unpolarized fragmentation function $D_{1}^{\Lambda/q}$ from the model scale $\mu_0$ to another scale through DGLAP evolution equation.
As for $D_{1T}^\perp$, we adopt the diagonal piece of the DGLAP evoluion kernel corresponding to twist-3 collinear correlation function $\hat{D}_{1T}^{\perp(3)}(z,z,\mu_b)$ from Ref.~\cite{Kang:2010xv}, which has the same form as that for the unpolarized fragmentation function.

When applying the model results for $\hat{D}_{1T}^{\perp(3)}$ and $D_1$ as the nonperturbative input to calculating the $\Lambda$ polarization, some attention should be paid, since the results are obtained as a low scale $\mu_0^2=0.23$ GeV$^2$, which is beyond the limit of applicability of the perturbative QCD evolution equations. This is the common feature of many model calculations for TMD distributions and fragmentation functions because these models are usually valid at a hadronic scale.
As the evolutions for $\hat{D}_{1T}^{\perp(3)}$ and $D_1$ in the nonperturbative region are unknown, we assume they follow the same evolutions as those in the perturbative region.
This is of course an approximation to the real situation.
However, we expect the assumption will not qualitative change our result since $P_{\Lambda T}$ is the ratio between the polarized cross section and the unpolarized cross section, so that there is some cancelation of the effects between the numerator and the denominator.

\begin{figure}
  \centering
  % Requires \usepackage{graphicx}
  \includegraphics[width=0.32\columnwidth]{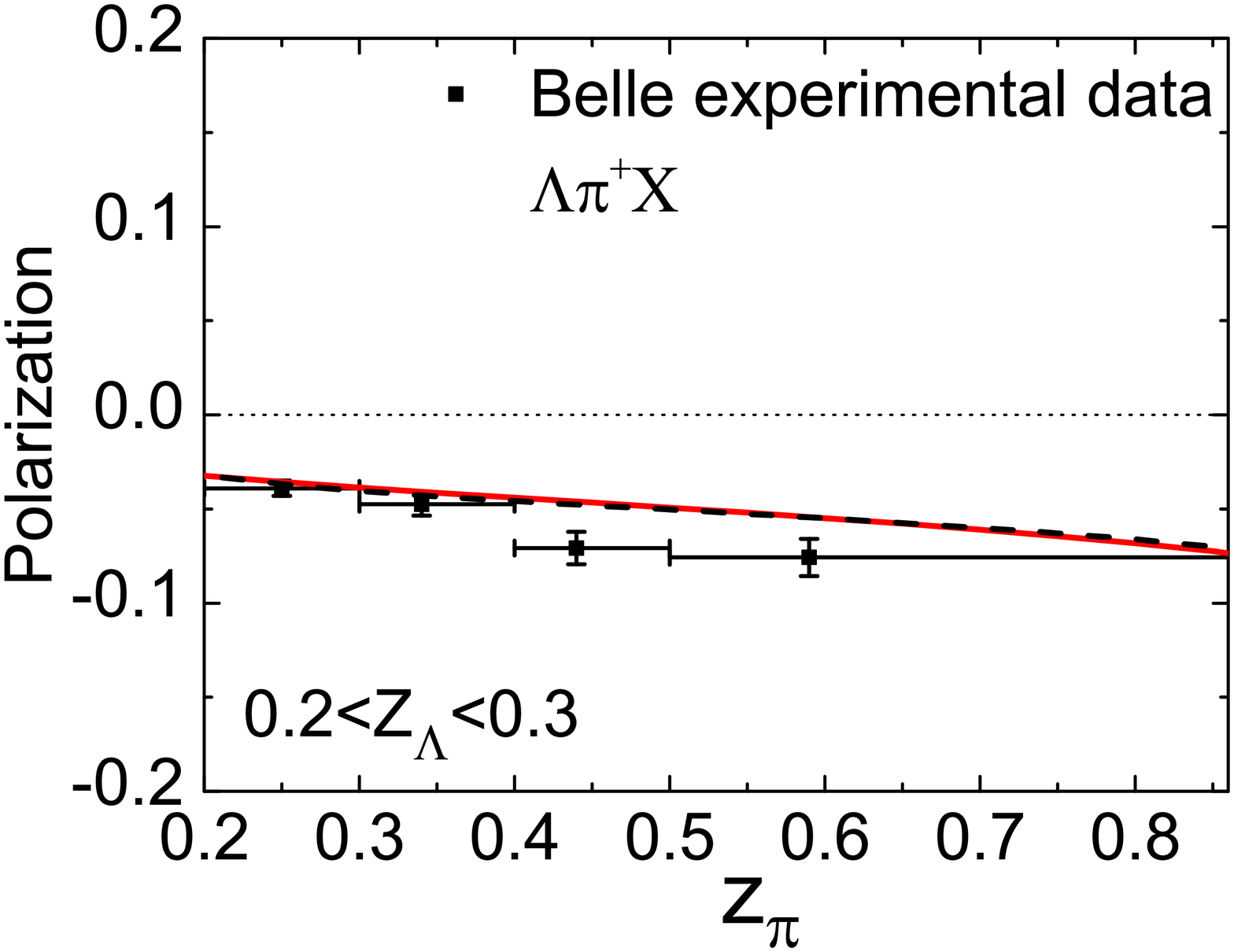}
  \includegraphics[width=0.32\columnwidth]{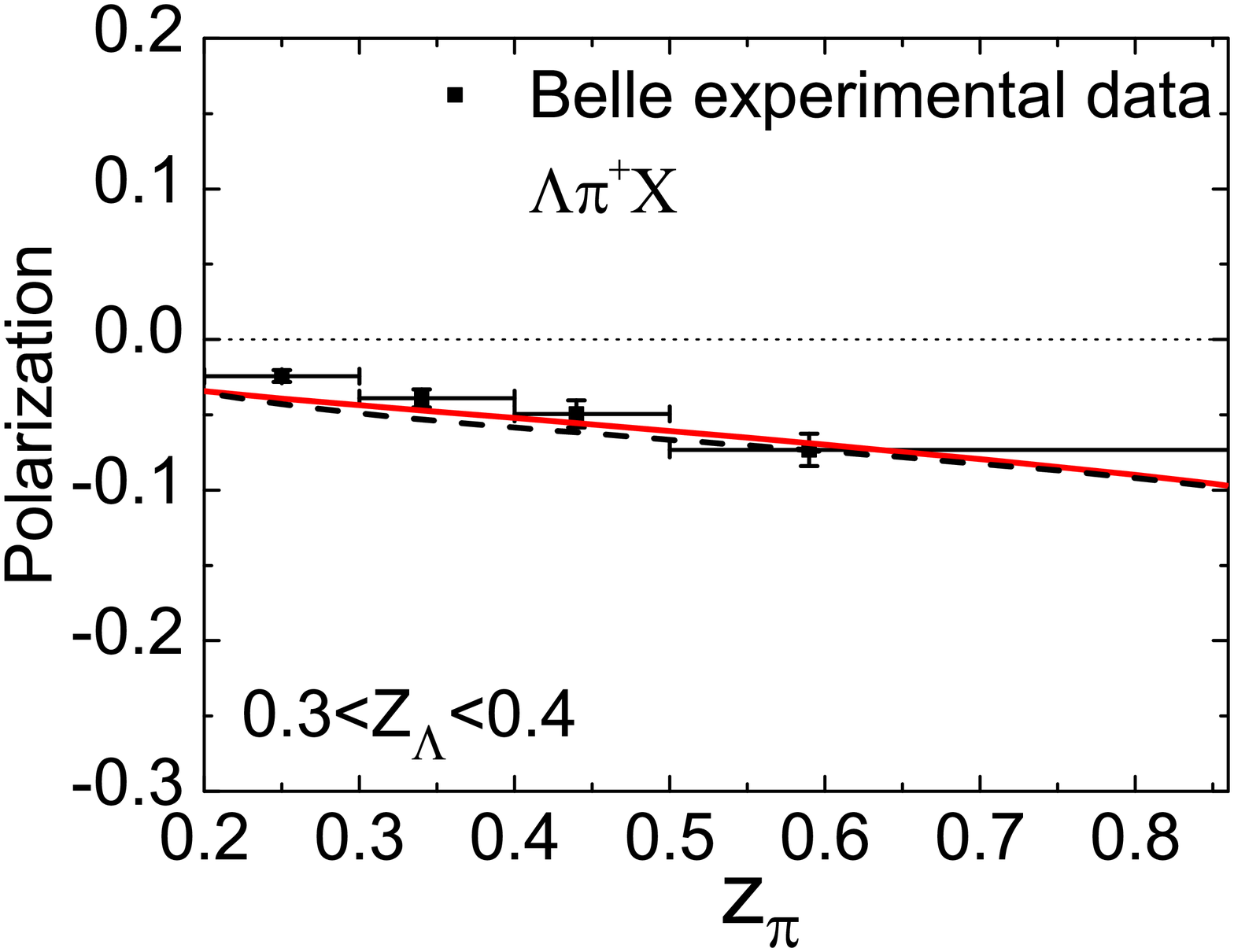}
  \includegraphics[width=0.32\columnwidth]{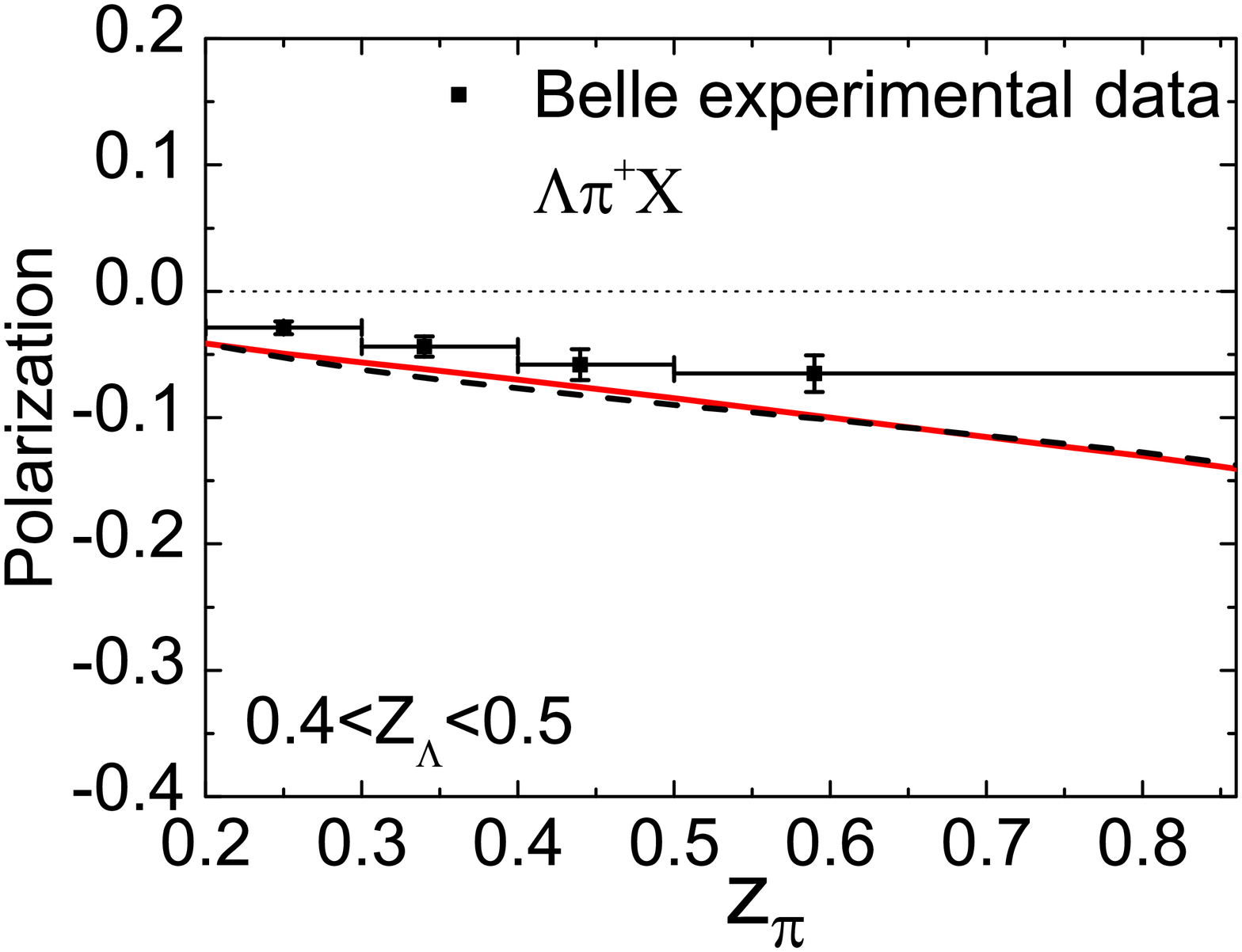}
  \caption{The $\Lambda$ transverse polarization as function of $z_\pi$ calculated in the process $e^+e^-\rightarrow\Lambda^\uparrow+\pi^{+}+X$. The red solid lines correspond to the results from the BDPRS parametrization~\cite{Bacchetta:2017gcc} [Eqs.~(\ref{eq:SNP-jhep1}) and Eq.~(\ref{eq:SNP-jhep2})] on the nonperturbative form factor while the dashed lines correspond to the results calculated from the EIKV parametrization~\cite{Echevarria:2014xaa} [Eqs.~(\ref{eq:SNP(unp-ff)})] on the nonperturbative Sudakov form factor. The solid squares represent the Belle data for comparison.
  }
  \label{fig1:pola-pi+}
\end{figure}

Note that the $z$-variable defined in the Belle experiments is the energy fraction of the hadron with respect to the fragmenting quark, denoted by $z_h={2E_h/\sqrt{s}}$ ($h= \Lambda$, $\pi$ or $K$), while our analysis in the previous sections and the input collinear function $\hat{D}_{1T}^{\perp(3)}(z,z,\mu_b)$ apply the light-cone momentum fraction of the hadron for $z$.
The two kinds of variables are connected by
\begin{align}
z_h= z\left[1+m_h^2/(z^2 s) \right] \, , \label{eq:z}
\end{align}
To achieve a better comparison between our calculation and the Belle data, we apply the transformation in Eq. (\ref{eq:z}) to change the light-cone momentum fraction $z$ to the energy fraction $z_h$ and calculate the polarization $P_{\Lambda T}$.
The Belle Collaboration measured the data in $z_\Lambda$ bins with boundaries at $z_{\Lambda }=0.2,0.3,0.4,0.5,0.9$~\cite{Guan:2018ckx}.
In Fig.~\ref{fig1:pola-pi+}, we plot the numerical results of the transverse polarization in the process $e^+e^-\rightarrow \Lambda^\uparrow \pi^+ X$ at $Q=10.58$ GeV as functions of $z_\pi$ for different $z_\Lambda$ bins.
Since the positivity bound of $D_{1T}^\perp$ in the model calculation was violated at large $z$ region ($z>0.75$)~\cite{Yang:2017cwi}, the $z_{\Lambda}$ bins are shown in Fig.~\ref{fig1:pola-pi+} up to 0.5, i.e. the analysis is performed in $z_1$ bins with boundaries at $z_{\Lambda}=0.2,0.3,0.4,0.5$.
To make the TMD factorization valid in the kinematic region $P_{h\perp}/z_h\ll Q$, the integration over the transverse momentum $P_{h\perp}$ is performed in the region of $0<P_{h\perp}<0.7$ GeV.
The solid lines show the $\Lambda$ transverse polarization calculated from the BDPRS evolution formalism~\cite{Bacchetta:2017gcc} and the $b_\ast$ prescription in Eq.~(\ref{b*2}), while the dashed lines show the $\Lambda$ transverse polarization using the EIKV parametrization~\cite{Echevarria:2014xaa} for the nonperturbative Sudakov form factor for comparison, with the $b_\ast$ prescription in Eq.~(\ref{eq:b*}).
The solid squares show the experimental data measured by the Belle Collaboration~\cite{Guan:2018ckx}, with the error bars including both the systematic error and the statistical error.

\begin{figure}
  \centering
  % Requires \usepackage{graphicx}
  \includegraphics[width=0.32\columnwidth]{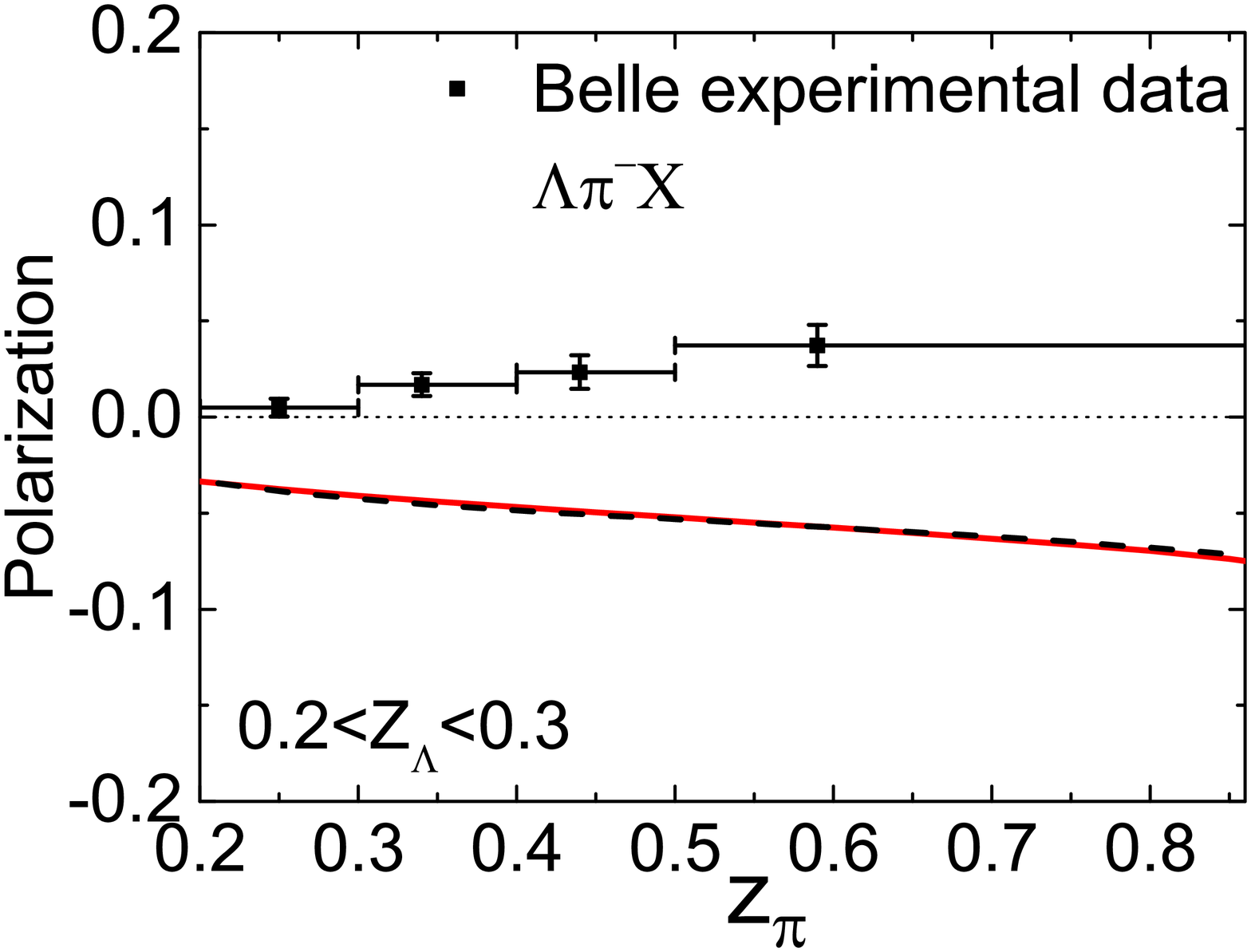}
  \includegraphics[width=0.32\columnwidth]{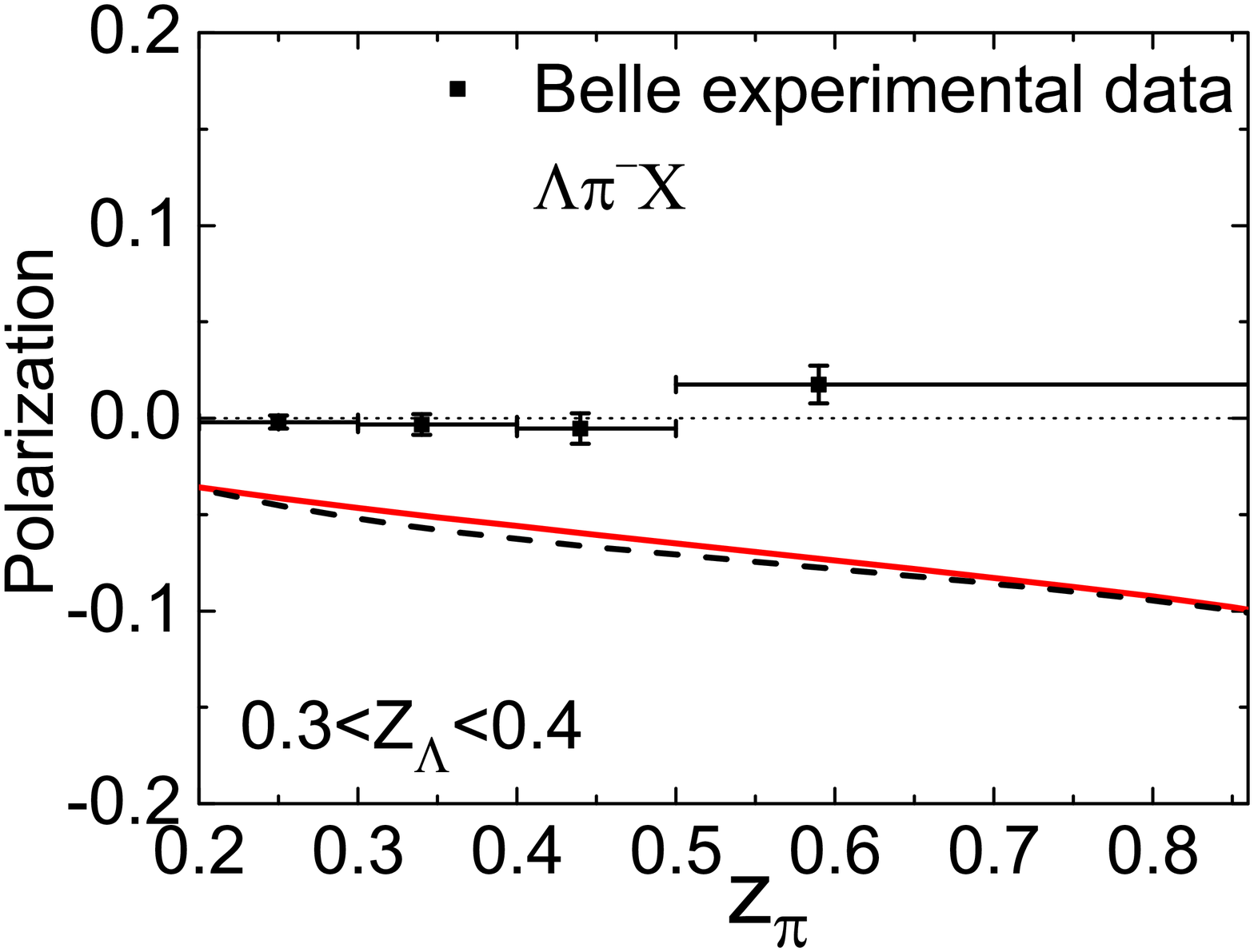}
  \includegraphics[width=0.32\columnwidth]{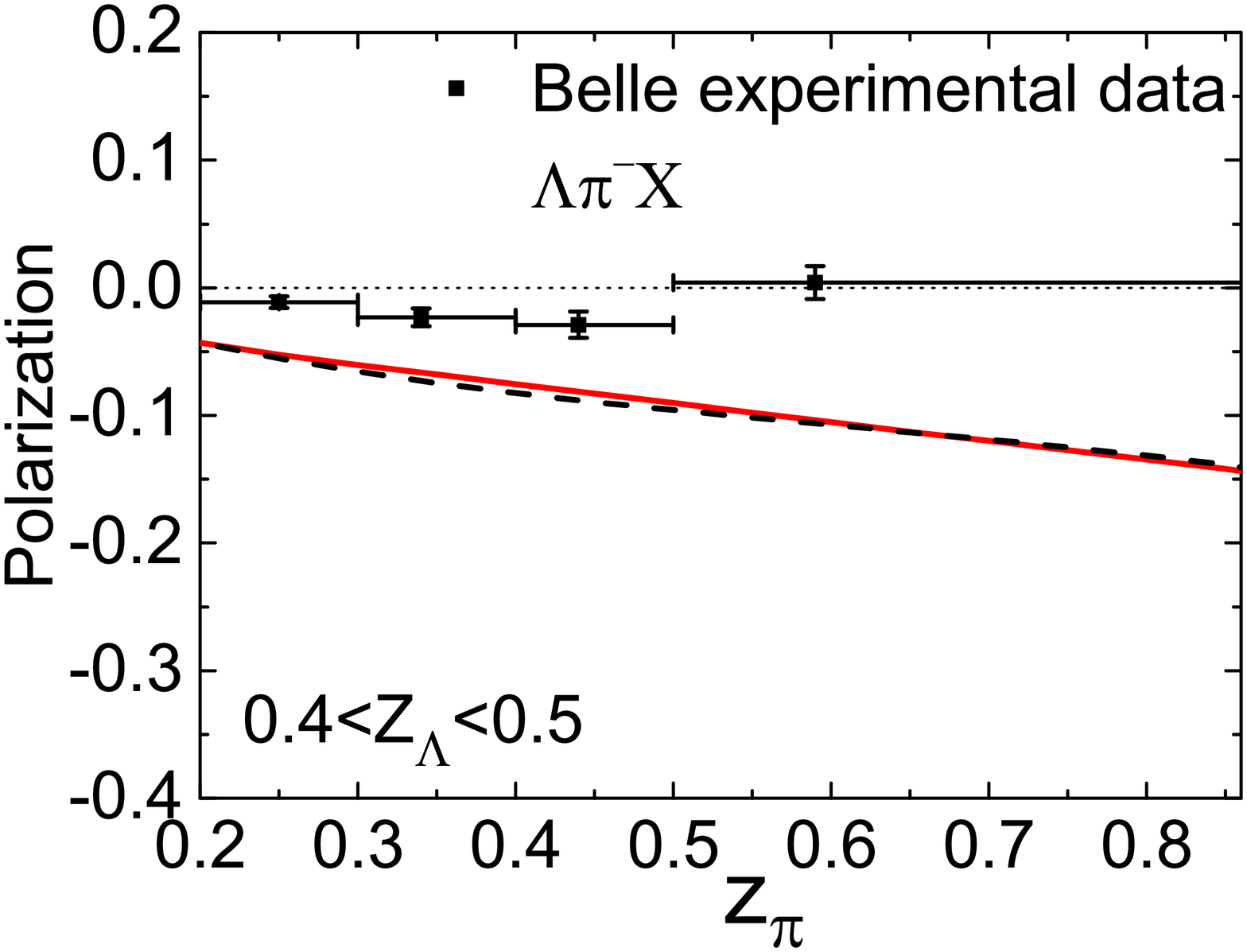}
  \caption{Similar to Fig.~\ref{fig1:pola-pi+}, but the $\Lambda$ transverse polarization in the process $e^+e^-\rightarrow\Lambda^\uparrow+\pi^{-}+X$.}
  \label{fig1:pola-pi-}
\end{figure}

\begin{figure}
  \centering
  % Requires \usepackage{graphicx}
  \includegraphics[width=0.32\columnwidth]{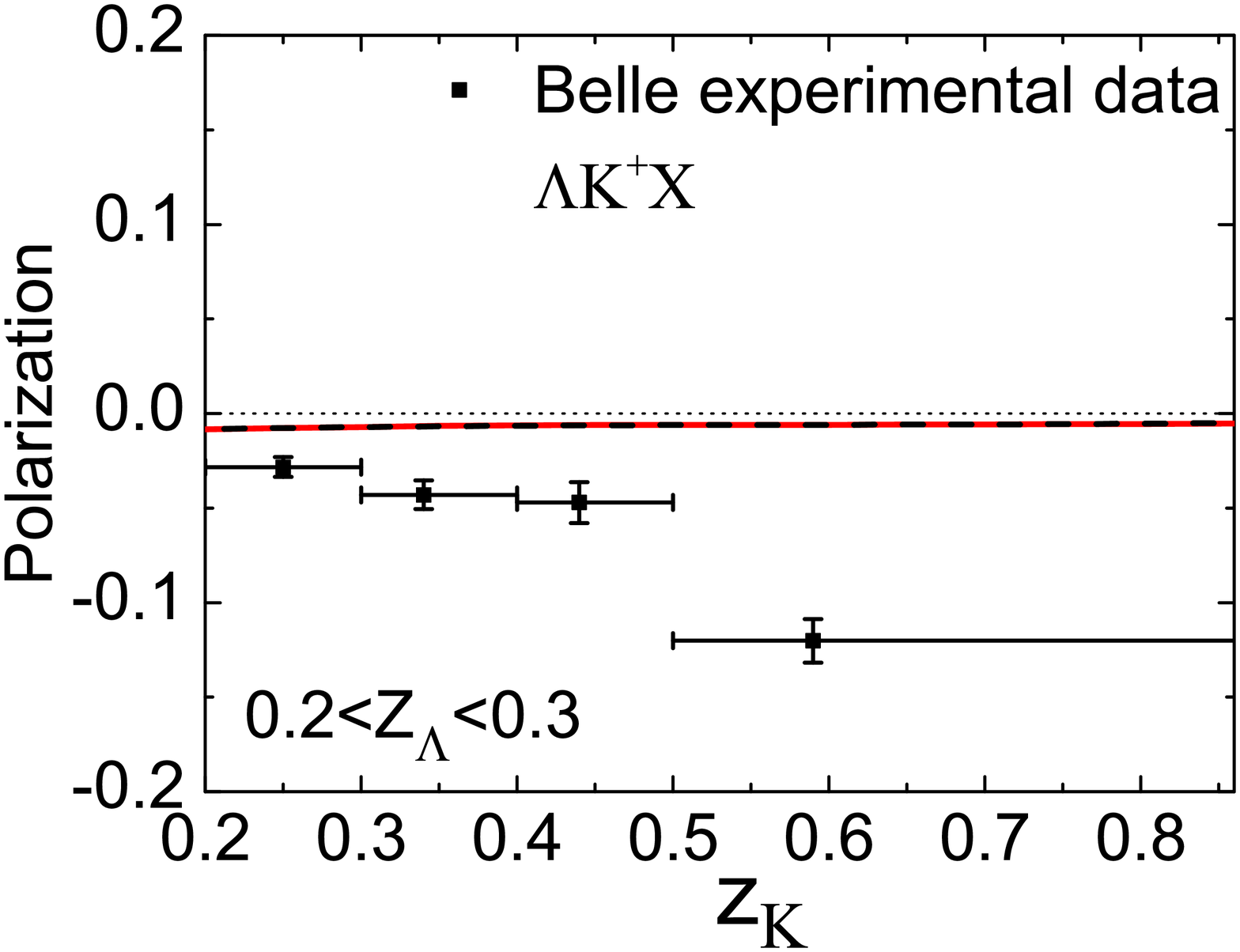}
  \includegraphics[width=0.32\columnwidth]{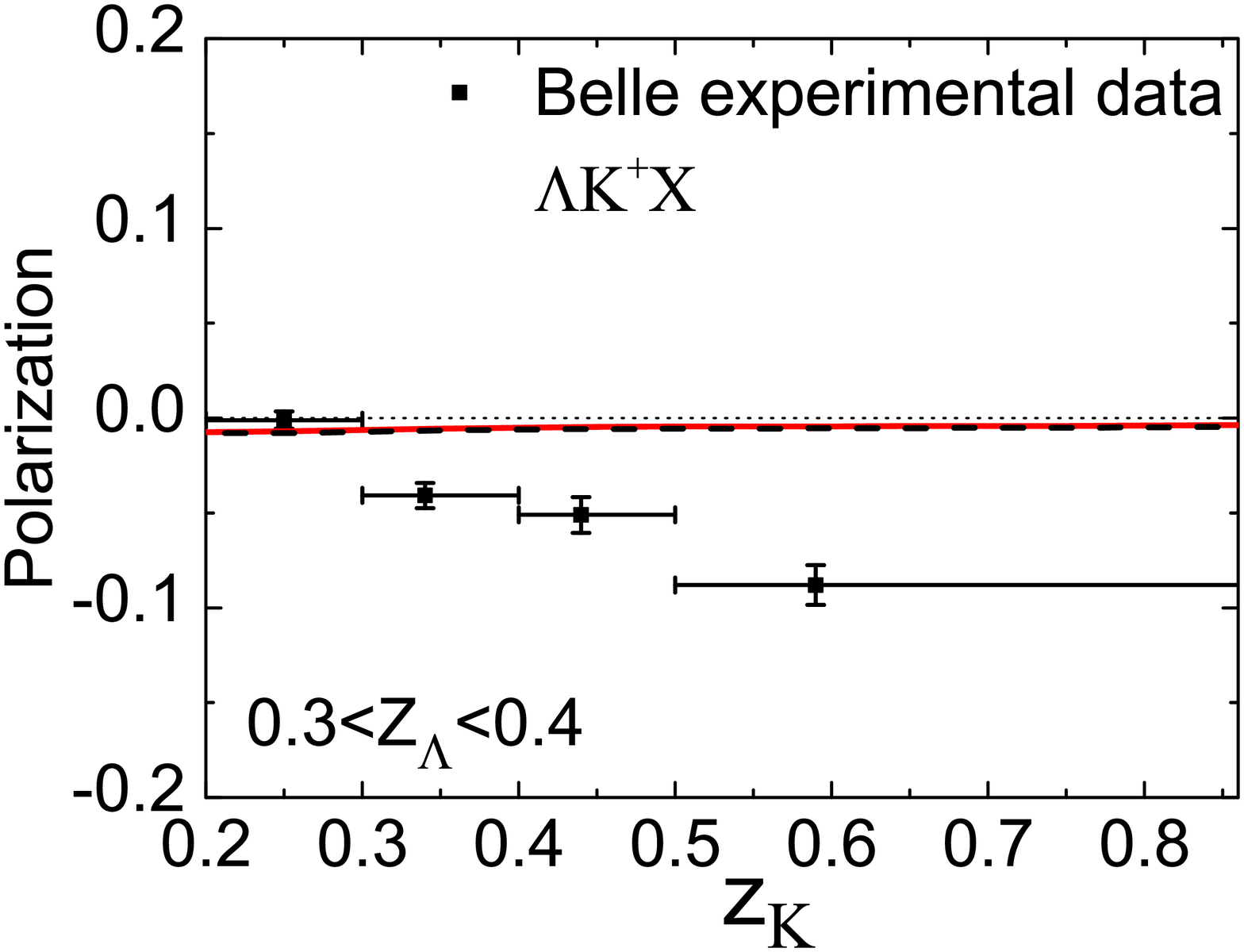}
  \includegraphics[width=0.32\columnwidth]{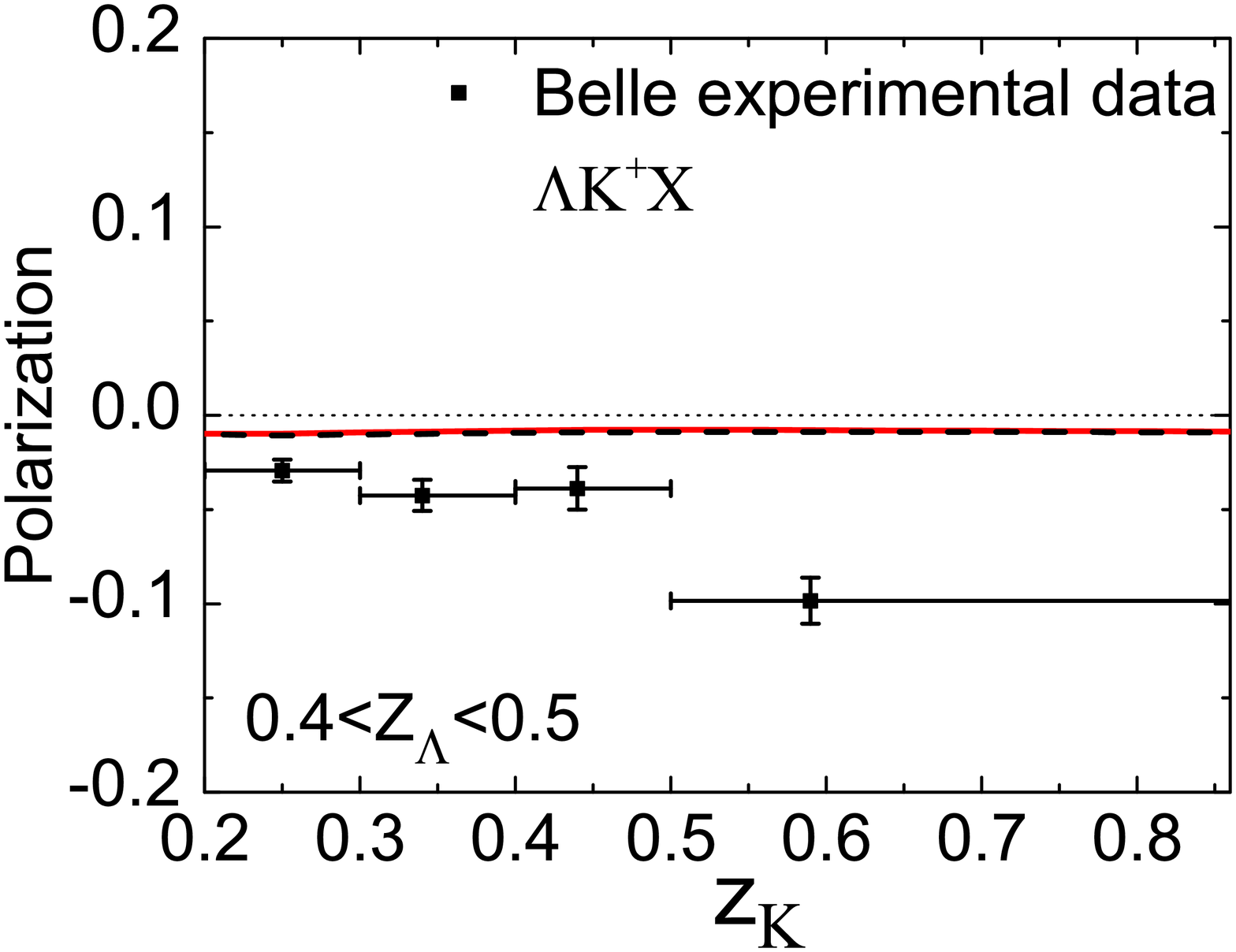}
  \caption{Similar to Fig.~\ref{fig1:pola-pi+}, but the $\Lambda$ transverse polarization calculated in the process $e^+e^-\rightarrow\Lambda^\uparrow+K^{+}+X$.}
  \label{fig2:pola-k+}
\end{figure}

\begin{figure}
  \centering
  % Requires \usepackage{graphicx}
  \includegraphics[width=0.32\columnwidth]{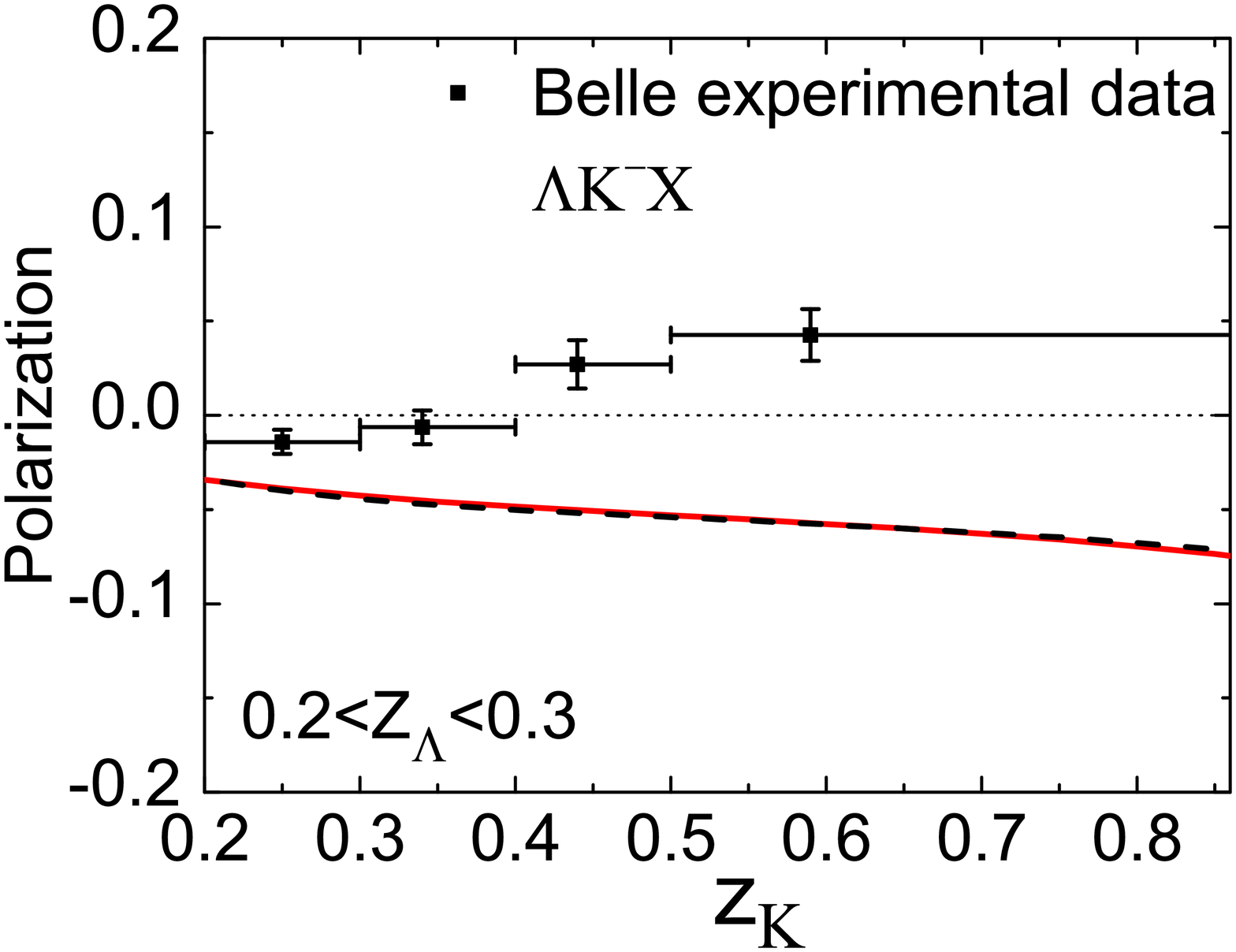}
  \includegraphics[width=0.32\columnwidth]{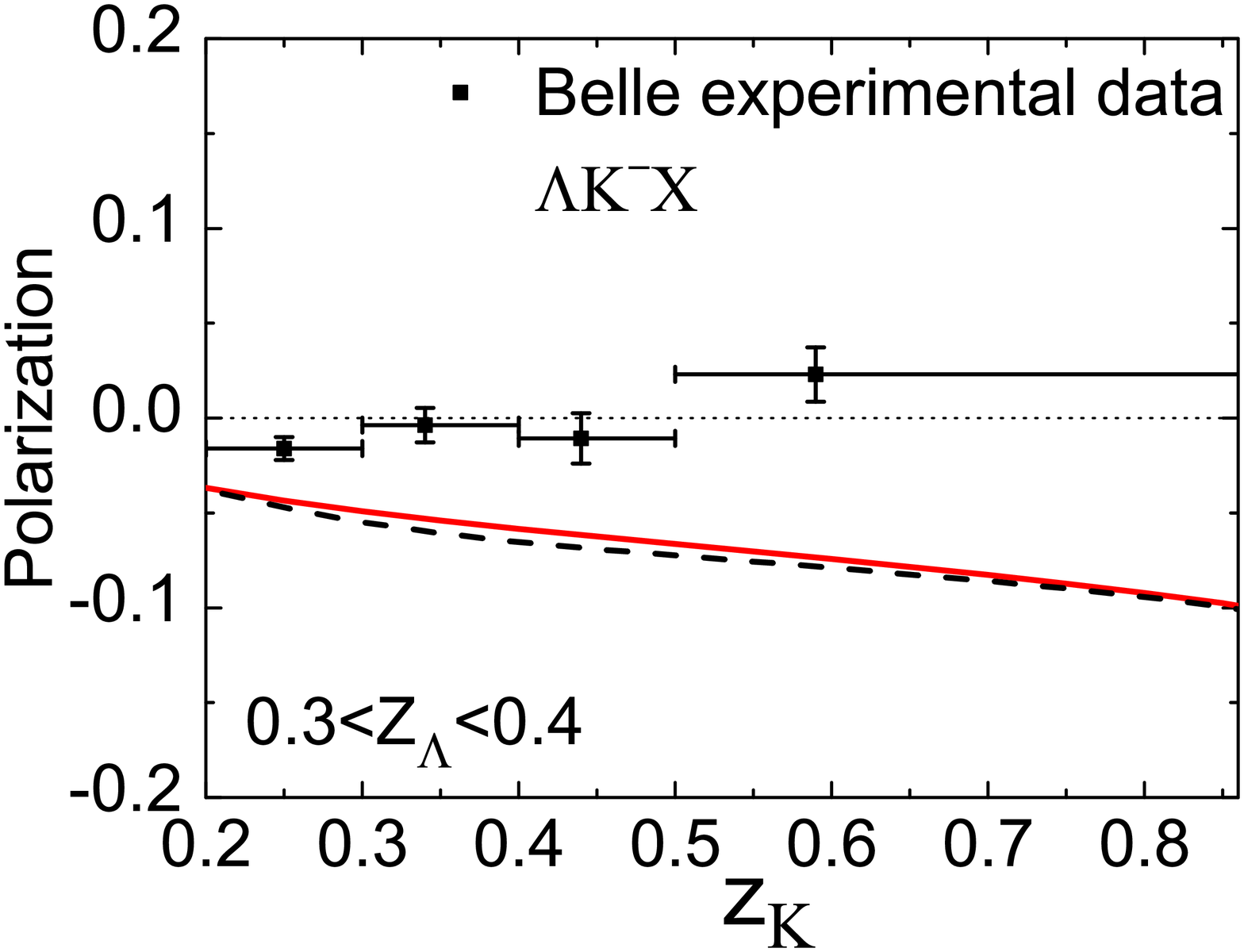}
  \includegraphics[width=0.32\columnwidth]{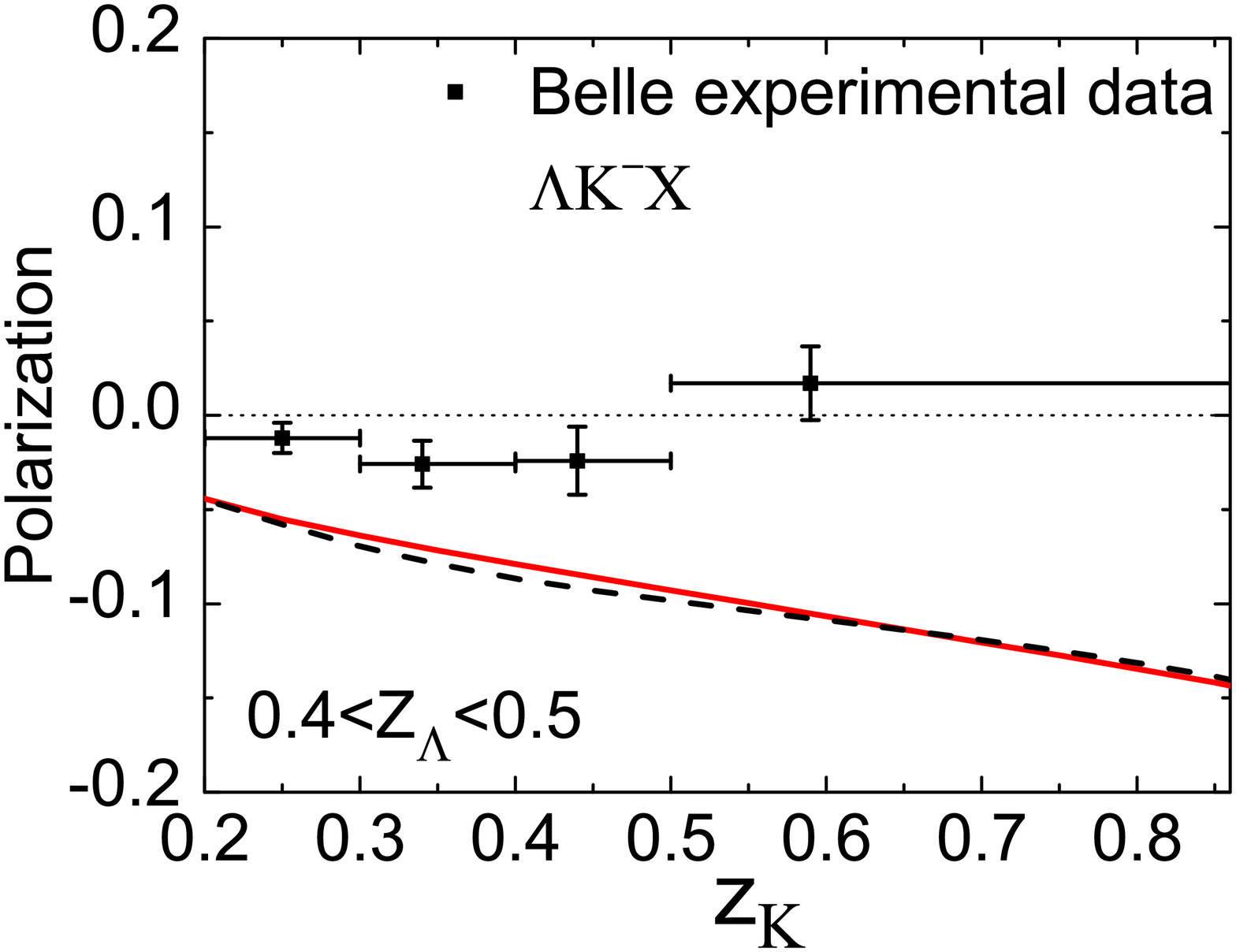}
  \caption{Similar to Fig.~\ref{fig1:pola-pi+}, but the $\Lambda$ transverse polarization calculated in the process $e^+e^-\rightarrow\Lambda^\uparrow+K^{-}+X$.}
  \label{fig3:pola-k-}
\end{figure}

As shown in Fig.~\ref{fig1:pola-pi+}, the transverse polarization of $\Lambda$ in the process $e^+e^-\rightarrow \Lambda^\uparrow \pi^+ X$ is negative, in agreement with most of the Belle data.
In this case, the polarization is dominated by the term $D_{1T}^{\perp\Lambda/d}\otimes D_1^{\pi^+/\bar{d}}$, as in our model there is no sea contribution to fragmentation functions and $D_{1T}^{\perp\Lambda/s}\ll D_{1T}^{\Lambda/d}$.
Thus, the fact that $D_{1T}^{\perp\Lambda/d}$ is negative in our model lead to the prediction of negative polarization in $\Lambda \pi^+$ production .
Our estimates also show that the size of the polarization increases with increasing $z_\pi$, and the results from the EIKV parameterization are close to those from the BDPRS parametrization.
We can conclude that different methods dealing with the non-perturbative evolution lead to the similar results for the polarization $P_{\Lambda T}$.
We also find that using different definition for $z$ (momentum fraction or energy fraction) will slightly reduce the size of the numerical result.
In our calculation, using the energy fraction $z_h$ can describe the Belle data better than using the momentum fraction $z$.

In Fig.~\ref{fig1:pola-pi-}, we plot the the transverse polarization of $\Lambda$ vs $z_\pi$ in different $z_\Lambda$ bins, but in the process $e^+e^-\rightarrow \Lambda^\uparrow \pi^- X$.
In this case the polarization from our model calculation is firmly negative in each $z_\Lambda$ bin. The reason is that in $\Lambda^\uparrow \pi^-$ production our model indicates the term $D_{1T}^{\perp\Lambda/u}\otimes D_1^{\pi^-/\bar{u}}$ dominates.
This result is similar to our model prediction for the polarization in $\Lambda \pi^+$ production, since in our model $ D_{1T}^{\perp u}= D_{1T}^{\perp d}$ and $ D_{1T}^{\perp s}$ is much smaller than $ D_{1T}^{\perp u}$, as shown in Fig.~\ref{fig0}.
However, the polarization from the Belle data for $\Lambda^\uparrow \pi^- $ production is positive in the $0.2<z_\Lambda<0.3$ bin, and turn to be negative in the $0.4<z_\Lambda<0.5$ region with relatively smaller size.
In the intermediate bin $0.3<z_\Lambda<0.4$, the polarization is consistent with zero.
This shows that our calculation for $\Lambda^\uparrow \pi^-$ production strongly disagrees with the Belle data.
In addition, the charge-conjugation symmetry implies $P_{\Lambda T}(\bar{\Lambda}h^-)=P_{\Lambda T}(\Lambda h^+)$. In these respects, the calculated results for $P_{\Lambda T}(\bar{\Lambda}\pi^-)$ are consistent with experimental data of Belle.

Finally, we also calculate the transverse polarization of $\Lambda$ in the processes $e^+e^-\rightarrow \Lambda^\uparrow K^+ X$ and $e^+e^-\rightarrow \Lambda^\uparrow K^- X$.
The results are shown in Fig.~\ref{fig2:pola-k+} and Fig.~\ref{fig3:pola-k-}, respectively.
We conclude that the overall signs of the polarization in both the $\Lambda^\uparrow K^+$ and $\Lambda^\uparrow K^-$ productions are negative, consistently with the sign of the Belle data for the $\Lambda^\uparrow K^+$ production and the $\Lambda^\uparrow K^-$ production in the lower $z_K$ region. However, our predicted polarization in the $\Lambda^\uparrow K^+$ production is rather small, which is significantly different from the Belle data; while our prediction in the $\Lambda^\uparrow K^-$ production is several times larger than the Belle data.
The reason for a poor model result in $\Lambda^\uparrow K^+$ production is because the $D_{1T}^{\perp\Lambda/s}\otimes D_1^{K^+/s}$ term provides the main contributions, while in our model $D_{1T}^{\perp\Lambda/s}$ is much smaller than that of the $u$ and $d$ quarks.
In the case of $\Lambda^\uparrow K^-$ production, the dominant contribution is the $D_{1T}^{\perp \Lambda/u}\otimes D_1^{K^-/\bar{u}}$ term, leading to a rather large polarization.
In addition, the charge-conjugation symmetry implies that $P_{\Lambda T}(\bar{\Lambda} K^-) = P_\Lambda(\Lambda K^+)$ and $P_{\Lambda T}(\bar{\Lambda} K^+) =P_{\Lambda T}(\Lambda K^-)$.

Some comments are in order.
Although our prediction for the polarization in $\Lambda^\uparrow \pi^+$ production agrees with the Belle data, there are large discrepancies between the Belle data and our predictions for $\Lambda^\uparrow \pi^-$ production as well as $\Lambda^\uparrow K^{\pm}$ production.
From the available parameterizations on $D_{1T}^\perp$ from Refs.~\cite{DAlesio:2020wjq} and \cite{Callos:2020qtu} which can satisfactorily describe the polarizations in $\Lambda^\uparrow \pi^\pm$
and $\Lambda^\uparrow K^\pm$ productions, one can find that a positive $D_{1T}^{\perp \Lambda/u}$, a negative $D_{1T}^{\perp \Lambda/d}$ and substantial sea contribution are crucial for describing the Belle data in all cases.
For example, positive polarizations in $\Lambda^\uparrow \pi^-$ production requires a positive $D_{1T}^{\perp \Lambda/u}$; a negative sea PFF $D_{1T}^{\perp \Lambda/\bar{u}}$ will lead to a negative nonzero polarization in $\Lambda^\uparrow K^+$~\cite{DAlesio:2020wjq,Callos:2020qtu}.
These features are different from our model input, in which $D_{1T}^{\perp \Lambda/u}$ is negative and is equal to $D_{1T}^{\perp \Lambda/d}$.
The origin of $D_{1T}^{\perp \Lambda/u}=D_{1T}^{\perp \Lambda/d}$ comes from the assumption of $SU(6)$ symmetry in the model calculation of the $\Lambda$ fragmentation function.
Besides, the spectator model does not incorporate sea contents of the $\Lambda$ fragmentation functions.
For a further study, a model that can allocate substantial SU(6) breaking effect may lead to very different result for the PFFs of the u and d quarks, particularly in sign.
This will improve the model description of the polarization in $\Lambda^\uparrow \pi^-$ production at Belle.
Incorporating the sea quarks in the model could also provide a better description on the $\Lambda^\uparrow K^\pm$ production.
Finally, in our calculation we have not considered the $Y$-term in the cross section.
The inclusion of the $Y$-term could change the size of the Polarization in the large momentum region.

\section{Conclusion}
\label{Sec.conclusion}
The $e^+e^-$ annihilation with a transversely polarized $\Lambda$ hyperon produced in the final state is a useful tool to study the hyperon spin structure and the non-perturbative fragmentation mechanism.
In this work, we have studied the transverse polarization of $\Lambda$ in the processes  $e^+e^-\rightarrow\Lambda(\bar{\Lambda})+\pi^{\pm}+X$ and $e^+e^-\rightarrow\Lambda(\bar{\Lambda})+K^{\pm}+X$ at Belle by applying the TMD factorization formalism.
The polarization was contributed by the convolution of the T-odd PFF $D_{1T}^\perp$, which describes the fragmentation of an unpolarized quark into a transversely $\Lambda$, and the unpolarized fragmentation function of pion or kaon.
To evolve the TMD fragmentations of the pion, kaon and $\Lambda$ from the initial energy to the experimental energy, we have taken into account the TMD evolution effects, in which the Sudakov form factor plays an important role.
The Sudakov form factor is separated into perturbative part and non-perturbative part, for the former one, we adopted the results from the perturbative QCD at NLL accuracy, while for the latter one, we considered two different non-perturbative TMD evolution formalism for comparison.
As the nonperturbative Sudakov form factor associated with the $\Lambda$ PFF is still unknown, we assume that it has the same form as that of the unpolarized fragmentation function.
The hard coefficients associated with the corresponding collinear functions in the TMD evolution formalism are kept at the leading-order accuracy.
For the fragmentation functions of $\Lambda$, we have chosen the results from the diquark spectator model.

It has been shown that different choices of nonperturbative Sudakov form factors in the TMD evolution formalism lead to similar results for transverse polarization of $\Lambda$ in process $e^+e^-$ annihilation.
Within the framework of TMD evolution, our prediction for the polarization in $\Lambda^\uparrow \pi^+$ production agrees with the Belle data.
However, there are large discrepancies between the Belle data and our predictions for $\Lambda^\uparrow \pi^-$ production as well as $\Lambda^\uparrow K^{\pm}$ production.
A comparison with our model input for the PFF with the available parameterizations Refs.~\cite{DAlesio:2020wjq} and \cite{Callos:2020qtu} indicates that a positive $D_{1T}^{\perp \Lambda/u}$ and sizable sea PFF are essential to describe the many aspects of the Belle data.
Possible improvements of the model calculation include allocating substantial SU(6) breaking effect to lead to a opposite sign of the u quark PFF  compared with the d quark PFF, incorporating the sea quarks in the model, and taking into the $Y$-term in the large momentum region.
Further studies are needed to understand the transverse polarization of the Lambda production in $e^+ e^-$ annihilation from the model aspect.

\section*{Acknowledgements}

This work is partially supported by the National Science Foundation of China (grants No. 11575043,11905187,11847217). X. Wang is supported by China Postdoctoral Science Foundation under Grant No.~2018M640680 and the Academic Improvement Project of Zhengzhou University.
Y. Yang is supported by Shandong Provincial Natural Science Foundation, China (Grants No. ZR2020QA081).

\end{document}